\documentclass[useAMS]{mn2e}
\newcommand{\gtappeq}{\raisebox{-0.6ex}{$\;\stackrel{\raisebox{-.2ex}{$\textstyle >$}}{\sim}\;$}}
\newcommand{\ltappeq}{\raisebox{-0.6ex}{$\;\stackrel{\raisebox{-.2ex}{$\textstyle <$}}{\sim}\;$}}

\usepackage{times}
\usepackage{graphicx}

\usepackage{psfig} 

\title[The Donor Stars of Cataclysmic Variables]{The Donor Stars of
Cataclysmic Variables} 
\author[Christian Knigge]{Christian Knigge$^1$\\
School of Physics \& Astronomy, University of Southampton,
Highfield, Southampton SO17 1BJ, UK, christian@astro.soton.ac.uk} 
\begin{document}

\date{Accepted ????. Received ????}

\pagerange{\pageref{firstpage}--\pageref{lastpage}} \pubyear{2006}

\maketitle

\label{firstpage}

\begin{abstract}

We carefully consider observational and theoretical constraints on the
global properties of secondary stars in cataclysmic variable stars
(CVs). We then use these constraints to construct and test a complete,
semi-empirical donor sequence for CVs with orbital periods $P_{orb}
\leq 6$~hrs. All key physical and photometric parameters of CV
secondaries (along with their spectral types) are given as a function
of $P_{orb}$ along this sequence. This provides a benchmark for
observational and theoretical studies of CV donors and evolution. 

The main observational basis for our donor sequence is an empirical
mass-radius relationship for CV 
secondaries. Patterson and co-workers have recently shown that this 
can be derived from superhumping and/or eclipsing CVs. We
independently revisit all of the key steps in this derivation,
including the calibration of the period
excess-mass ratio relation for superhumpers and the use of a single
representative primary mass for most CVs. We also present an optimal
techique for estimating the parameters of the mass-radius relation
that simultaneously ensures consistency with the observed locations of
the period gap and the period minimum. We 
present new determinations of these periods, finding $P_{gap,+} = 3.18
\pm 0.04$~hrs (upper edge), $P_{gap,-} = 2.15 \pm 0.03$~hrs (lower edge)
and $P_{min} = 76.2 \pm 1.0$~min (period minimum).

We test the donor sequence by comparing observed and predicted
spectral types ($SpTs$) as a function of orbital period. To this end,
we update the $SpT$ compilation of Beuermann and  co-workers and
show explicitly that CV donors have later $SpTs$ than main sequence (MS)
stars at all orbital periods. This extends the conclusion of the
earlier study to the short-period regime ($P_{orb} < 3$~hrs). We then 
compare our donor sequence to the CV data, and find that it does an 
excellent job of matching the observed $SpTs$. Thus the empirical 
mass-radius relation yields just the right amount of radius expansion
to account for the later-than-MS spectral types of CV donors. There
is remarkably little intrinsic scatter in both the mass-radius and
$SpT-P_{orb}$  relations, which confirms that most CVs follow a unique
evolution track.

The donor sequence exhibits a fairly sharp drop in temperature,
luminosity, and optical/infrared flux well before the minimum
period. This may help to explain why the detection of brown dwarf
secondaries in CVs has proven to be extremely difficult.

We finally apply the donor sequence to the problem of distance
estimation. Based on a sample of 22 CVs with  
trigonometric parallaxes and reliable 2MASS data, we show that
the donor sequence correctly traces the envelope of the observed
$M_{JHK}-P_{orb}$ distribution. Thus robust lower limits on distances
can be obtained from single-epoch infrared observations. However, for
our sample, these limits are typically about a factor of two below the
true distances.

\end{abstract}


\section{Introduction}
\label{intro}

Cataclysmic variable stars (CVs) are compact, interacting binary
systems in which a white dwarf primary accretes from a low-mass,
roughly main-sequence donor star. The mass transfer and secular
evolution of CVs is driven by angular momentum losses. In systems
with long orbital 
periods ($P_{orb} \gtappeq 3$~hrs), the dominant angular momentum loss
mechanism is thought to be magnetic braking (MB) due to a stellar wind
from the donor star. In the canonical ``disrupted magnetic braking''
evolution scenario for CVs, MB stops when the secondary becomes fully convective,
at $P_{orb} \simeq 3$~hrs. At this point, the donor detaches from the
Roche lobe, and gravitational radiation (GR) becomes the only
remaining angular momentum loss mechanism. The GR-driven shrinkage of
the orbit ultimately brings the secondary back into contact at $P_{orb}
\simeq 2$~hrs, at which point mass transfer resumes. The motivation
for this scenario is the dearth of mass-transferring CVs in the
so-called period gap between $P_{orb} \simeq 2$~hrs and $P_{orb} \simeq
3$~hrs.  

As a CV evolves, its donor star is continuously losing mass. As long
as the mass-loss time scale ($\tau_{\dot{M}_2} \sim M_2/\dot{M}_2$) is
much longer than the donor's thermal time  scale ($\tau_{th} \sim
\frac{GM^2_2}{R_2 L_2}$), the secondary is able to maintain thermal
equilibrium and should closely follow the standard main sequence track
defined by single stars. Here and throughout, we use $M_2$, $R_2$ and
$L_2$ to denote a donor's mass, radius and luminosity, and $\dot{M}_2$
to denote the rate at which it is losing mass to the primary. If the
condition $\tau_{\dot{M}_2} >> \tau_{\dot{M}_{th}}$ is not met, the
donor will be driven out of thermal equilibrium and become oversized
compared to an isolated main sequence (MS) star of identical mass. 

So how do thermal and mass-loss time scales compare for the donor
stars in CVs? Or, to put it another way, are the donors main
sequence stars? This question has been addressed twice in recent 
years, in very different ways. First, Beuermann et al. (1998; 
hereafter B98) showed that, at least for periods $P_{orb} 
\gtappeq 3$~hrs (i.e. above the period gap), the spectral types
($SpT$s) of CV donor stars are significantly later than those of
isolated MS stars.
\footnote{For the record, earlier studies along these lines were 
carried out by  Echevarr\'{i}a (1983), Patterson (1984), Friend et al
(1990ab) and Smith \& Dhillon (1998).} 
At the very longest periods ($P_{orb} \gtappeq
5-6$~hrs), this is probably due to the donors being somewhat evolved
(B98; also see Podsiadlowski, Han \& Rappaport 2003). However, for all
other systems, the late $SpT$s of the donors are simply a sign of
their losing battle to maintain thermal equilibrium.

Second, Patterson et al. (2005; hereafter P05) produced an empirical
mass-radius sequence for CV donors based on a sample of masses and
radii derived mainly from superhumping CVs. One of their key results
was that CV  donors are indeed oversized compared to MS stars, for all 
periods $P_{orb} \ltappeq 6$~hrs. They also detected a clear
discontinuity in $R_2$ for systems with similar $M_2$ above and below
the period gap. This is just what is expected in the disrupted
magnetic braking picture: the MB-driven systems just above the gap
should be losing mass faster than than the GR-driven systems just
below. They should therefore be further out of thermal equilibrium. 

In the present paper, we will build on these important studies
by constructing a complete, semi-empirical donor sequence for
CVs. More specifically, our goal is to derive a benchmark evolution
track for CV secondaries that incorporates all of the best existing
observational and theoretical constraints on the global donor properties.
\footnote{Here and below, we use the term ``donor sequence'' to
describe the dependence of the secondary star parameters on the
orbital period of a CV. We call a sequence ``complete'' 
if all important physical ($P_{orb}$, $M_2$,  $R_2$,
$T_{eff}$, $SpT$) and photometric (absolute magnitudes
in UBVRIJHKLM) donor properties are fully specified along the entire
evolutionary track.}
Along the way, we will update both of the earlier studies and show
that they are mutually consistent: the mass-radius relation obtained
by P05 is just what is needed to account for the late $SpT$s 
observed by B98. 

Our donor sequence should be useful for many practical
applications. Perhaps most fundamentally, it provides a 
benchmark for what we mean by a ``normal'' CV and can thus be used to 
test CV evolution scenarios that predict the properties of CV
donors. The sequence can also be used to simplify and improve on
``Bailey's method'' for estimating the distances to CVs 
(Bailey 1981). The new version of the method requires only knowledge
of the orbital period 
and an infrared magnitude measurement. Conversely, for systems with
known distances, the sequence can be used to estimate the donor
contribution to the system's flux in any desired bandpass. Finally,
the present work is also a stepping stone towards another goal: the
construction of new, semi-empirical mass-transfer rate and angular
momentum loss laws for CVs. This should be possible, since the degree of
departure of a donor from the MS track is a direct measure of the mass
loss from it and hence of the angular momentum loss that drives
$\dot{M}_2$. 

\section{The empirical mass-radius relation for CV donor stars}

Given the importance of the $M_2-R_2$ relation to the present work, we
begin by revisiting and updating P05's $M_2$-$R_2$ relation for
CV donor stars. We still use the same fundamental data as P05, but
carry out a fully independent analysis to derive our own mass-radius
relation from the data. In the following sections, we will briefly
discuss each of the key steps in the derivation (and also introduce
some new ideas of our own). However, we start with an outline of the
overall framework of the method, i.e. the way in which a mass-radius
relation for CV secondaries can be derived from (mainly) observations
of superhumps.

\subsection{Donor Masses and Radii from Superhumping CVs}
\label{basics}

Superhumps are a manifestation of a donor-induced accretion disk
eccentricity that is observed primarily in erupting dwarf novae, but 
also in some non-magnetic nova-like CVs and even in a few LMXBs. Once
established, the eccentricity precesses on a time scale that is much
longer than the orbital period. The superhump signal is then observed
at the beat period between the orbital and precession periods. The
superhump period, $P_{sh}$, is therefore typically a few percent
longer than the orbital period, and the superhump excess, $\epsilon$,
is defined as  
\begin{equation}
\epsilon = \frac{P_{sh} - P_{orb}}{P_{orb}}.
\label{eps1}
\end{equation}

Both theory and observation agree that $\epsilon$ is a function of
the mass ratio, $q = M_2/M_1$. This $\epsilon-q$ relation can be
calibrated by considering eclipsing superhumpers for which an
independent mass ratio estimate is available. Given such a
calibration, we can estimate $q$ for any superhumper with measured
$\epsilon$. 

Two additional steps are needed to turn an estimate of the mass ratio
into an estimate of $M_2$ and $R_2$. First, in order to
obtain $M_2$ from $q$, we clearly first need an estimate of $M_1$. For
a few systems (mostly eclipsers), this can be measured 
directly. However, for most other systems, a representative value
has to be assumed. For example, P05 used $M_1 = 0.75 M_{\odot}$ for
all systems without a direct estimate, based on several 
estimates of the mean WD mass in CVs in the literature. With this
assumption, the mass of the donor is then simply given by  
\begin{equation}
M_2 = q M_1.
\label{q}
\end{equation}

The second additional step is to make use of the fact that the
secondary is filling its Roche lobe. The donor must therefore obey the
well-known period-density relation for Roche-lobe filling objects
(Warner 1995, Equation 2.3b)
\begin{equation}
<\rho_2> = 107 \; P^{-2}_{orb,h} ~ \rm{g\;cm^3}.
\label{dense}
\end{equation}
where $P_{orb,h}$ is the orbital period in units of hours and 
\begin{equation}
<\rho_2> = \frac{3 M_2}{4 \pi R_2^3}.
\label{dense2}
\end{equation}
Equation~\ref{dense} is accurate to better than 3\% over the interval
$0.01 < q < 1$. Since $M_2$ and $P_{orb}$ are known at this point, the
period-density relation can be recast to yield an estimate of the
donor radius
\begin{equation}
R_2/R_{\odot} = 0.2361 \;P^{2/3}_{orb,h} \;q^{1/3} \;(M_1/M_{\odot})^{1/3} 
\label{r2}
\end{equation}
or, equivalently, 
\begin{equation}
R_2/R_{\odot} = 0.2361 \; P^{2/3}_{orb,h} \; (M_2/M_{\odot})^{1/3}.
\label{r2_2}
\end{equation}
At this point, both $M_2$ and $R_2$ have been estimated. A fit to the
resulting set of $M_2$ and $R_2$ pairs (supplemented with additional
data points derived from from eclipsing systems) can then be used to
determine the functional form of the mass-radius relation for CV
donors. 

In the following sections, we will take a closer look at all of the key
steps we have outlined, and also add one final step of our own. More
specifically, we will consider (i) the calibration of the $\epsilon-q$ 
relation for superhumpers; (ii) the assumption of a single $M_1$ value
for most CVs; (iii) the external constraints that the final mass-radius
relation should reproduce the observed locations of the period gap and
the minimum period; (iv) the derivation of an optimal fit to the
$M_2$-$R_2$ pairs, allowing for correlated errors, intrinsic
dispersion and external constraints. 

\subsection{Calibrating the $\epsilon-q$ Relation}
\label{sec:calib}

Table~7 in P05 provides a list of calibrators for the $\epsilon-q$
relation. This contains 10 superhumping and 
eclipsing CVs with independent mass ratio constraints, one
superhumping CV with a large 
superhump excess and an {\em assumed} upper limit on $q$ (BB Dor), and 
also one superhumping and eclipsing LMXB with a very low mass ratio
(KV UMa). In devising our own, independent calibration of the
$\epsilon-q$ relation, we used the same set of calibrators, but
analysed them independently. Details are given in Appendix~A; the
results are shown in Figure~\ref{fig:epsq}. Our preferred fit to the
data is given by
\begin{equation}
q(\epsilon) = (0.114 \pm 0.005) + (3.97 \pm 0.41) \times (\epsilon - 0.025)
\label{eps3}
\end{equation}
The errors here are 1-$\sigma$ for both parameters jointly, and the
shift applied to $\epsilon$ ensures that the fit 
parameters (and their errors) are reasonably uncorrelated. The fit is
shown in Figure~\ref{fig:epsq} and achieves a statistically acceptable
$\chi^2_\nu = 1.03$ without the need to add any intrinsic dispersion
in excess of 
the statistical errors on the $\epsilon$ and $q$ estimates. Thus any
intrinsic scatter around the calibrating relation must be small
compared to the statistical errors on the data
points. Figure~\ref{fig:epsq} also 
allows a direct comparison of our fit to the data against P05's, as
well as against two recent theoretically-motivated calibrations of the
$\epsilon-q$ relationship (Goodchild \& Ogilvie 2006; Pearson
2006). All calibrations agree quite well, except at the highest mass
ratios, where data are sparse.

\begin{figure}
%\begin{minipage}{126mm}
\includegraphics[width=84mm]{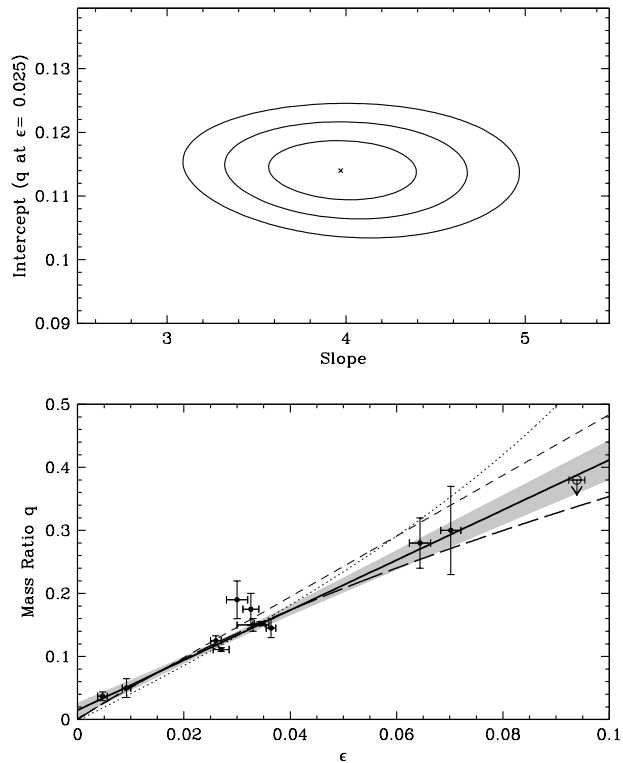}
\caption{Calibration of the $q-\epsilon$ relation for superhumping
CVs. {\em Top panel:} Constraints on the slope and intercept of the
preferred linear calibration. The cross marks the best-fitting
parameter combination. The ellipses correspond to 1$\sigma$, 2$\sigma$
and 3$\sigma$ contours on both parameters jointly. {\em Bottom panel:}
Mass ratio vs period excess for the calibrating stars. The thick solid
line is the best-fitting linear calibration, with the shaded area
marking the statistical 1-$\sigma$ range. The thick long-dashed line
shows P05's calibration, the thin dotted line is Pearson's (2006)
calibration, and the thin short-dashed line is the calibration of
Goodchild \& Ogilvie (2006). The data point shown as a thin open
symbol shows the upper limit on the mass ratio of BB Dor proposed by
P05.}
\label{fig:epsq}
%\end{minipage}
\end{figure}

One final point worth noting is that the statistical errors on the fit
parameters (and indeed the uncertainty regarding the functional form
of the calibration itself) translate 
into {\em systematic} errors on the resulting masses and radii. The
1-$\sigma$ error band arising from the statistical uncertainties on our
fit parameters is indicated by the dashed region in
Figure~\ref{fig:epsq}. Formally, 
this is less than 10\% even out to $q \simeq 0.4$, but the fit is very
poorly constrained beyond $q \simeq 
0.3$. The $\epsilon - q$ relation could thus change shape in this
regime. The {\em statistical} error on a mass ratio obtained via
Equation~\ref{eps3} can be estimated by folding the error on
$\epsilon$ through the calibrating relation. 

\subsection{The Assumption of Constant Primary Mass}
\label{primary}

The next key step in the derivation of masses and radii from 
superhumpers is the assumption of a single primary mass for most
systems in the sample. The worry here is not so much that the 
assumed mass might differ from the true mean WD mass for CVs. This 
would simply shift all ($\log{M_2}$, $\log{R_2}$) pairs along a line
of slope 1/3, but would not alter the {\em shape} of the donor mass-radius
relation. Instead, the main concern is that $M_1$ might exhibit
systematic trends within the observed CV population, most importantly
with orbital period. Such trends {\em could} affect the shape of the
mass-radius relation.

\begin{figure}
%\begin{minipage}{126mm}
\includegraphics[width=84mm]{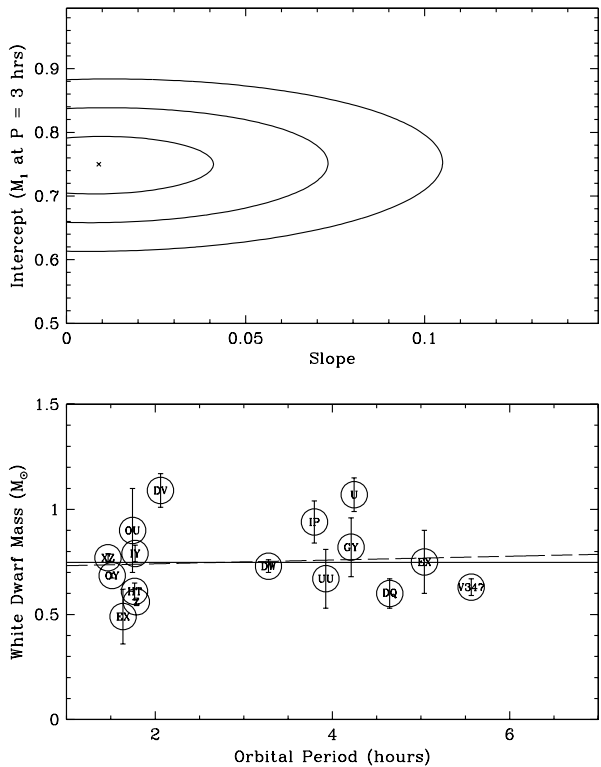}
\caption{Constraints on white dwarf mass evolution with orbital
period. {\em Top panel:} Results of a linear fit to $M_{WD}$ vs
$P_{orb}$ for eclipsing systems with well-determined white dwarf
mass. The cross marks the best-fitting parameter combination. The
ellipses correspond to 1$\sigma$, 2$\sigma$ and 3$\sigma$ contours on
both parameters jointly. {\em Bottom panel:} $M_{WD}$ vs $P_{orb}$ for
eclipsers. The thick solid line shows the optimally estimate mean
white dwarf mass for all systems: $M_{WD} = 0.75 \pm 0.05 M_{\odot}$
The thin dashed line shows the best linear fit to the data.}
\label{fig:m1}
%\end{minipage}
\end{figure}

In order to test whether this is a problem, we can check if
there is any dependence of $M_1$ on $P_{orb}$ among CVs with reliable
WD mass estimates (i.e. eclipsing systems). P05 actually tabulate all
available $M_1$ estimates derived from eclipsing CVs. In the bottom
panel of Figure~\ref{fig:m1}, we plot these estimates as a function of
$P_{orb}$ for all systems with $P_{orb} < 6$~hrs. We exclude systems with
longer periods, since they have evolved donors and thus follow a 
different evolution track (see Section~\ref{donors_vs_MS} below).

Figure~\ref{fig:m1} does not provide evidence for evolution of $M_1$ with
$P_{orb}$. A full discussion of the statistical evidence for this
assertion is given in Appendix~B, along with a discussion of how this
statement can be reconciled with earlier work that seemed to show 
such evolution. Here, we simply note that a linear fit to the data
yields a slope consistent with zero. The best-fitting 
constant $M_1$ yields a mean WD mass of $<M_1> = 0.75 \pm 0.05
M_{\odot}$ and requires an intrinsic scatter of $\sigma_{int} =
0.16 M_{\odot}$. Thus we confirm P05's assertion that, in the absence
of other information, $M_1$ can be estimated as $M_1 = 0.75 M_{\odot}$
with about 20\% uncertainty.

We finally note that the global intrinsic dispersion -- $\sigma_{M_1} =
0.16~M_{\odot}$, i.e. 21\% -- is the {\em statistical} error
associated with taking $M_1 = 0.75 M_{\odot}$ for any particular
system. By contrast, the formal error on the mean WD mass ($0.05
M_{\odot}$) translates into a {\em systematic} error on donor masses
and radii (since a shift in the assumed $M_1$ affects all affected
data points in the same way). 

Having dealt with the $\epsilon-q$ calibration and the assumption of 
constant $M_1$, we are ready to calculate an updated set of $M_2$ and
$R_2$ values (with statistical errors) for all superhumpers. We list
these  in Table~\ref{tab:superhumpers}. Our estimates are not  
identical to those derived by P05, since we have used a different
$\epsilon-q$ calibration and have explicitly accounted for the
intrinsic dispersion in $M_1$ when estimating the statistical
errors. However, as expected, the overall pattern presented by the
data is much the same. The main change in the data values is a
slight shift towards higher masses, which arises because our
$q$-estimates are generally a little higher than P05's for fixed
$\epsilon$ (see Figure~\ref{fig:epsq}).

\subsection{External Constraints: The Location of the Period Gap and 
the Minimum Period}     
\label{sec:constraints}

In principle, we could now simply fit the $M_2-R_2$ pairs in
Table~\ref{tab:superhumpers} (supplemented with data for eclipsing
systems). However, there are actually additional empirical constraints
that can (and should) be imposed on the mass-radius relation. These
constraint come from the observed locations of the period gap 
and of the period minimum. 

\begin{figure*}
\begin{minipage}{170mm}
\includegraphics[width=170mm]{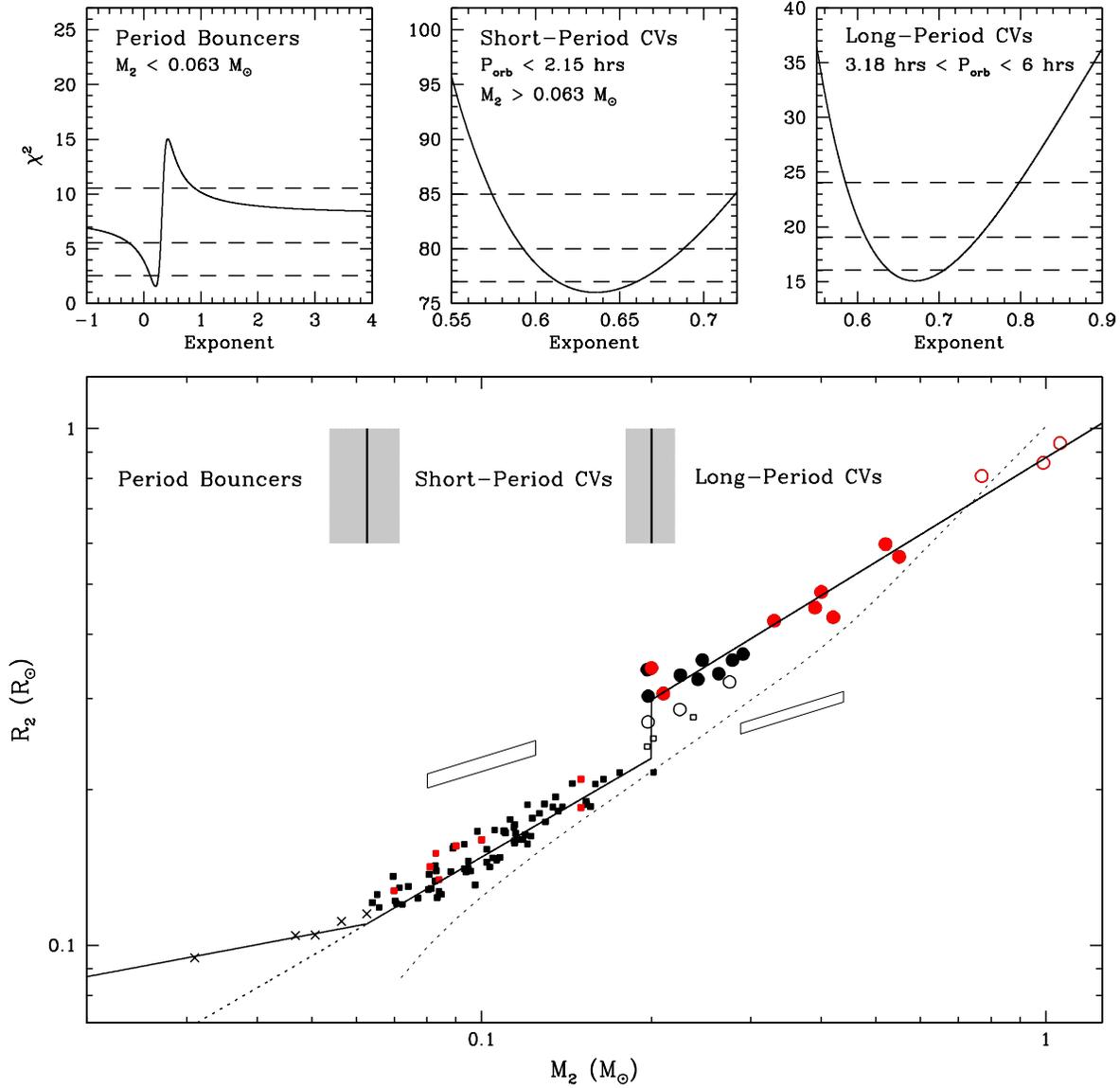}
\caption{
%\footnotesize
The mass-radius relation of CV donor stars. {\em Bottom
panel:} Points shown are empirical mass and radius estimates for CV
donors. Superhumpers are shown in black, eclipsers in red. Filled
squares correspond to short-period CVs, filled circles to long-period
systems, and crosses to likely period bouncers. The parallelograms
illustrate the typical error on a single short-period or long-period
CV. Open symbols were ignored in fits to the data since they
correspond to systems in the period gap or long-period (probably
evolved) systems. The solid lines show the optimal fit to the data in
the period bouncer, short-period and long-period regions. The dotted
line is the mass-radius relation for main sequence stars taken from
the 5~Gyr BCAH98 isochrone.
{\em Top panels:} Constraints on the power law exponents of the
$M_2-R_2$ relations in the three period/mass regimes. For each regime,
we plot $\chi^2$ vs exponent and indicate the $\chi^2$
corresponding to 1-$\sigma$, 2-$\sigma$ and 3-$\sigma$ around the
minimum with horizontal dashed lines. The strange shape of the
$\chi^2$ curve for the period bouncers near the exponent $\frac{1}{3}$
is real. It arises because the intrinsic power law relationship
between $M_2$ and $R_2$ estimates has the same exponent.
\vspace*{2cm}
}
\label{fig:m2r2}
\end{minipage}
\end{figure*}

Let us first consider the period gap. The bottom panel in
Figure~\ref{fig:m2r2} shows the $M_2-R_2$ estimates from
Table~\ref{tab:superhumpers}, along with similar data for 
eclipsing systems, taken from Table~8 in P05. There is a clear 
discontinuity in donor  radii at $M_2 \simeq 0.2 M_{\odot}$, with
donors in long-period systems being 
larger than those in short-period systems. As discussed in more detail
by P05, the transition is quite sharp and broadly consistent with the
standard CV evolution scenario. Briefly, we should expect systems just
above and below the period gap to have identical donor masses, since 
CVs evolve through the period gap as detached binaries, with no
significant mass loss from the secondary. However, their radii must 
differ, since both short- and long-period donors
must still obey the period-density relation for Roche-lobe-
filling stars (Equation~\ref{dense}). More specifically, if we denote
the upper and lower edges of the period gap as $P_{gap,\pm}$, the
ratio of donor radii at the gap edges must satisfy
\begin{equation}
\frac{R_{2,+}}{R_{2,-}} =
\left(\frac{P_{gap,+}}{P_{gap,-}}\right)^{2/3}.
\end{equation}
Physically, donors above the gap are larger since they lose mass at a
much higher rate than those below and are therefore forced more out of
thermal equilibrium.

If we accept the premise that CV donors evolve through the period gap
as detached systems, the data in Figure~\ref{fig:m2r2} suggests 
that the critical donor mass at which mass transfer stops 
(and then restarts) is $M_{conv} = 0.20 \pm 0.02 M_{\odot}$, where
the estimated error is purely statistical.
\footnote{We use the subscript {\em
conv} to denote this critical mass, since in the canonical picture it
corresponds to the mass at which the donor becomes fully convective.}
In other words, $M_{conv}$ is the donor mass at both edges of the
period gap. Given empirical estimates for the location of these
edges, we can therefore use the period-density relation to determine
$R_{2,conv}$, i.e. the radii of donors at the gap edges. Thus the
locations of the gap edges ($P_{gap,\pm}$) fix the $M_2-R_2$ relations
for long- and short-period systems at $M_{conv}$ and $P_{gap,\pm}$. 

\begin{figure}
%\begin{minipage}{126mm}
\includegraphics[width=84mm]{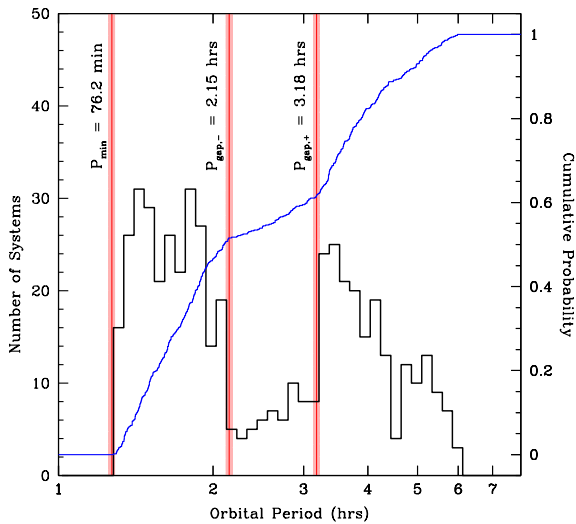}
\caption{Differential and cumulative orbital period distribution of
CVs, based on data taken from Edition 7.6 of the Ritter \& Kolb
(2003) catalog. Our adopted values for the minimum period and the
period gap edges are shown as vertical lines. The shaded regions
around them indicate our estimate of the errors on these values.}
%\end{minipage}
\label{fig:periods}
\end{figure}

Figure~\ref{fig:periods} shows the orbital period distribution of CVs
drawn from the Ritter \& Kolb (2003) catalog (Edition 7.6) in 
both differential and cumulative form. The period gap is obvious in these
distributions. We have carried out repeated measurements of the gap
edges from this data with a variety of binning schemes. Based on these
measurements, we estimate $P_{gap,-} = 2.15 \pm 0.03$~hrs and
$P_{gap,+} = 3.18 \pm 0.04$~hrs. It is worth noting that the number of
systems inside the gap increases towards longer periods. This may be
due to low-metallicity systems, which are expected to invade the gap
from above (Webbink \& Wickramasinghe 2002).

\begin{table*}
\footnotesize
\begin{minipage}{175mm}
%\begin{minipage}{200mm}
\caption{Donor masses and radii as estimated from superhump
periods. Orbital periods are in hours; masses and radii are in solar
units. Note that mass and radius estimates are not independent,
but correlated via Equation~\ref{r2_2}.}
\begin{tabular}{@{}lcccccclcccccclcccccclccccc}
\hline
System &
$P_{orb}$ &
$M_2$ &
$\sigma_{M_2}$ &
$R_2$ &
$\sigma_{R_2}$ &
\hspace{2.0cm} &
System &
$P_{orb}$ &
$M_2$ &
$\sigma_{M_2}$ &
$R_2$ &
$\sigma_{R_2}$ \\
\hline
      DI UMa   &    1.3094   &    0.051    &   0.011    &   0.105    &   0.008     &      &          CY UMa  &    1.6697   &    0.119  &     0.026    &   0.164    &   0.012 \\
    V844 Her   &    1.3114   &    0.083    &   0.018    &   0.124    &   0.009 	   &	  &          FO And  &    1.7186   &    0.115  &     0.027    &   0.165    &   0.013 \\
      LL And   &    1.3212   &    0.097    &   0.023    &   0.131    &   0.010 	   &	  &          OU Vir  &    1.7450   &    0.130  &     0.029    &   0.173    &   0.013 \\
SDSS 0137-09   &    1.3289   &    0.085    &   0.019    &   0.125    &   0.009 	   &	  &          VZ Pyx  &    1.7597   &    0.110  &     0.024    &   0.165    &   0.012 \\
ASAS 0025+12   &    1.3452   &    0.072    &   0.017    &   0.120    &   0.009 	   &	  &          CC Cnc  &    1.7645   &    0.156  &     0.034    &   0.186    &   0.014 \\
      AL Com   &    1.3601   &    0.047    &   0.010    &   0.104    &   0.008 	   &	  &          HT Cas  &    1.7676   &    0.089  &     0.009    &   0.154    &   0.005 \\
      WZ Sge   &    1.3606   &    0.056    &   0.009    &   0.111    &   0.006 	   &	  &          IY UMa  &    1.7738   &    0.093  &     0.006    &   0.157    &   0.003 \\
  RX 1839+26   &    1.3606   &    0.063    &   0.015    &   0.115    &   0.009 	   &	  &          VW Hyi  &    1.7825   &    0.110  &     0.024    &   0.166    &   0.012 \\
      PU CMa   &    1.3606   &    0.077    &   0.018    &   0.123    &   0.009 	   &	  &           Z Cha  &    1.7880   &    0.089  &     0.003    &   0.155    &   0.001 \\
      SW UMa   &    1.3634   &    0.084    &   0.020    &   0.127    &   0.010 	   &	  &          QW Ser  &    1.7887   &    0.110  &     0.026    &   0.167    &   0.013 \\
      HV Vir   &    1.3697   &    0.071    &   0.015    &   0.120    &   0.009 	   &	  &          WX Hyi  &    1.7954   &    0.114  &     0.025    &   0.169    &   0.012 \\
      MM Hya   &    1.3822   &    0.066    &   0.014    &   0.118    &   0.009 	   &	  &          BK Lyn  &    1.7995   &    0.154  &     0.033    &   0.187    &   0.013 \\
      WX Cet   &    1.3990   &    0.070    &   0.016    &   0.122    &   0.009 	   &	  &          RZ Leo  &    1.8250   &    0.114  &     0.026    &   0.171    &   0.013 \\
      KV Dra   &    1.4102   &    0.080    &   0.018    &   0.128    &   0.010 	   &	  &          AW Gem  &    1.8290   &    0.137  &     0.030    &   0.182    &   0.013 \\
       T Leo   &    1.4117   &    0.081    &   0.018    &   0.129    &   0.009 	   &	  &          SU UMa  &    1.8324   &    0.105  &     0.023    &   0.167    &   0.012 \\
      EG Cnc   &    1.4393   &    0.031    &   0.007    &   0.095    &   0.007 	   &	  &    SDSS 1730+62  &    1.8372   &    0.123  &     0.027    &   0.176    &   0.013 \\
   V1040 Cen   &    1.4467   &    0.103    &   0.023    &   0.142    &   0.011 	   &	  &          HS Vir  &    1.8456   &    0.153  &     0.033    &   0.190    &   0.014 \\
      RX Vol   &    1.4472   &    0.064    &   0.015    &   0.121    &   0.009 	   &	  &        V503 Cyg  &    1.8648   &    0.139  &     0.031    &   0.185    &   0.014 \\
      AQ Eri   &    1.4626   &    0.096    &   0.021    &   0.139    &   0.010 	   &	  &        V359 Cen  &    1.8696   &    0.127  &     0.030    &   0.180    &   0.014 \\
      XZ Eri   &    1.4678   &    0.094    &   0.005    &   0.139    &   0.003 	   &	  &          CU Vel  &    1.8840   &    0.098  &     0.024    &   0.166    &   0.013 \\
      CP Pup   &    1.4748   &    0.083    &   0.015    &   0.133    &   0.008 	   &	  &        NSV 9923  &    1.8984   &    0.134  &     0.030    &   0.185    &   0.014 \\
   V1159 Ori   &    1.4923   &    0.106    &   0.023    &   0.146    &   0.010 	   &	  &          BR Lup  &    1.9080   &    0.112  &     0.027    &   0.175    &   0.014 \\
   V2051 Oph   &    1.4983   &    0.095    &   0.022    &   0.141    &   0.011 	   &	  &       V1974 Cyg  &    1.9502   &    0.202  &     0.033    &   0.216    &   0.012 \\
    V436 Cen   &    1.5000   &    0.074    &   0.018    &   0.130    &   0.011 	   &	  &          TU Crt  &    1.9702   &    0.129  &     0.028    &   0.188    &   0.014 \\
      BC UMa   &    1.5026   &    0.102    &   0.022    &   0.145    &   0.010 	   &	  &          TY PsA  &    2.0194   &    0.135  &     0.030    &   0.194    &   0.014 \\
      HO Del   &    1.5038   &    0.093    &   0.022    &   0.141    &   0.011 	   &	  &          KK Tel  &    2.0287   &    0.121  &     0.027    &   0.187    &   0.014 \\
      EK TrA   &    1.5091   &    0.107    &   0.024    &   0.147    &   0.011 	   &	  &        V452 Cas  &    2.0304   &    0.159  &     0.035    &   0.205    &   0.015 \\
      TV Crv   &    1.5096   &    0.108    &   0.025    &   0.148    &   0.011 	   &	  &          DV UMa  &    2.0604   &    0.165  &     0.013    &   0.209    &   0.006 \\
      VY Aqr   &    1.5142   &    0.072    &   0.016    &   0.129    &   0.010 	   &	  &          YZ Cnc  &    2.0832   &    0.176  &     0.038    &   0.216    &   0.016 \\
      OY Car   &    1.5149   &    0.065    &   0.004    &   0.125    &   0.003 	   &	  &          GX Cas  &    2.1365   &    0.145  &     0.032    &   0.206    &   0.015 \\
  RX 1131+43   &    1.5194   &    0.088    &   0.019    &   0.139    &   0.010 	   &	  &          NY Ser  &    2.3460   &    0.197  &     0.043    &   0.242    &   0.018 \\
      ER UMa   &    1.5278   &    0.105    &   0.023    &   0.148    &   0.011 	   &	  &        V348 Pup  &    2.4442   &    0.202  &     0.045    &   0.251    &   0.019 \\
      DM Lyr   &    1.5710   &    0.095    &   0.022    &   0.145    &   0.011 	   &	  &        V795 Her  &    2.5982   &    0.237  &     0.051    &   0.276    &   0.020 \\
      UV Per   &    1.5574   &    0.081    &   0.019    &   0.137    &   0.010 	   &	  &        V592 Cas  &    2.7614   &    0.197  &     0.042    &   0.270    &   0.019 \\
      AK Cnc   &    1.5624   &    0.121    &   0.028    &   0.157    &   0.012 	   &	  &          TU Men  &    2.8128   &    0.225  &     0.049    &   0.286    &   0.021 \\
      AO Oct   &    1.5737   &    0.083    &   0.021    &   0.139    &   0.012 	   &	  &          AH Men  &    3.0530   &    0.275  &     0.059    &   0.323    &   0.023 \\
      SX LMi   &    1.6121   &    0.114    &   0.026    &   0.158    &   0.012 	   &	  &          DW UMa  &    3.2786   &    0.197  &     0.010    &   0.303    &   0.005 \\
      SS UMi   &    1.6267   &    0.118    &   0.026    &   0.160    &   0.012 	   &	  &          TT Ari  &    3.3012   &    0.263  &     0.056    &   0.335    &   0.024 \\
      KS UMa   &    1.6310   &    0.083    &   0.020    &   0.143    &   0.011 	   &	  &        V603 Aql  &    3.3144   &    0.242  &     0.044    &   0.327    &   0.020 \\
   V1208 Tau   &    1.6344   &    0.122    &   0.027    &   0.163    &   0.012 	   &	  &          PX And  &    3.5124   &    0.278  &     0.060    &   0.356    &   0.025 \\
      RZ Sge   &    1.6387   &    0.102    &   0.023    &   0.153    &   0.012 	   &	  &        V533 Her  &    3.5352   &    0.225  &     0.048    &   0.333    &   0.024 \\
      TY Psc   &    1.6399   &    0.114    &   0.025    &   0.159    &   0.012 	   &	  &          BB Dor  &    3.5808   &    0.291  &     0.062    &   0.366    &   0.026 \\
      IR Gem   &    1.6416   &    0.116    &   0.032    &   0.160    &   0.015 	   &	  &          BH Lyn  &    3.7380   &    0.246  &     0.053    &   0.356    &   0.026 \\
    V699 Oph   &    1.6536   &    0.070    &   0.017    &   0.136    &   0.011 	   &	  &          UU Aqr  &    3.9259   &    0.197  &     0.041    &   0.342    &   0.024 \\
\hline
\end{tabular}
\label{tab:superhumpers}
\end{minipage}
\end{table*}

%\clearpage

We can similarly demand that our mass-radius relationship should
reproduce the observed minimum period, $P_{min}$, of the CV 
population. Based again on the data in Figure~\ref{fig:periods}, we
estimate this to be $P_{min} = 76.2 \pm 1.0$~min. We can implement this
constraint by truncating the  $M_2-R_2$ relation for short-period CVs
at $M_{bounce}$, the donor mass where it reaches $P_{min}$. If there
are data points with lower masses, they need to be fit separately,
subject to the constraint that the fit should meet the mass-radius
relation for short-period CVs at $M_{bounce}$. Moreover, if
$P_{min}$ is supposed to be a {\em minimum} period, then the fit to
systems with $M_2 < M_{bounce}$ should yield an $M_2-R_2$ relation
that corresponds to {\em increasing} orbital period with decreasing
$M_2$. We will find below that that is indeed the case.

\subsection{The Optimal $M_2$-$R_2$ Relation for CV Donors}
\label{sec:fitting}

Given our set of mass-radius pairs and the added external constraints,
how can we obtain an optimal analytical description of the overall
$M_2-R_2$ relation? This question is not as trivial as it may seem,
for two reasons. First, the external constraints (i.e. the measured
values of $P_{min}$, $P_{gap,\pm}$ and $M_{conv}$) need to be
imposed self-consistently on fits to the $M_2-R_2$ pairs. Second, the
mass and radius estimates for any given superhumper (along with their
errors) are completely correlated. This was already pointed out
by P05 and is easy to see from Equation~\ref{r2_2}, which shows that
$R_2 \propto M_2^{1/3}$. The masses and radii of eclipsers are
similarly correlated, since only  
$q$ and $M_1$ are generally estimated directly from the light curve,
with $M_2$ and $R_2$ being obtained indirectly in much the same way as
for superhumpers.\footnote{We actually calculated our own $R_2$
estimates and errors for eclipsers from the $M_1$, $M_2$ and
$P_{orb}$ values listed in Table~8 of P05; this ensures consistency,
i.e. all of the data satisfy the same period-density relation. As
expected, our numbers agreed very well with those in P05's Table~8. }
This correlation should be taken into account when fitting the data.  

In Appendix~C, we show how the standard $\chi^2$-statistic can be
modified so as to explicitly account for correlated data points,
external constraints and intrinsic dispersion. When carrying out our
$\chi^2$ fits, we explicitly distinguish between three groups of CVs,
each of  
which is fit independently with a power-law $M_2-R_2$ relation: (i)
non-evolved long-period CVs ($3.18$~hrs~$< P_{orb} < 6$~hrs); (ii)
``normal'' short-period CVs ($P_{orb} < 2.15~hrs$ and $M_2 >
M_{bounce}$; (iii) period bouncers ($M_2 < M_{bounce}$). The
value of $M_{bounce}$ is obtained from the fit to short-period CV,
as described in Section~\ref{sec:constraints}. In principle, this
could be an iterative procedure, since the value of $M_{bounce}$
determines which points are included in the fit. In practice, the fit
to the short-period systems is quite insensitive to the exact value,
since the vast majority of points lie well above any plausible
estimate for $M_{bounce}$. For the optimal fit described below, we
find $M_{bounce} = 0.0626 M_{\odot}$.

Remarkably, we find that the intrinsic radius scatter required by the 
$\chi^2$ fits to the short- and long-period systems is only about
2\%-3\%. For comparison, the typical statistical error is about
7\%-8\%. This implies that the intrinsic dispersion is probably rather
poorly constrained by the data, but also that it cannot be more than a
few percent. Thus there really is a unique evolution track that is  
followed by almost all CVs. When other tracers are used to study CV 
evolution (e.g. luminosity or period changes), the existence of this
track is often masked by variability on time scales shorter than the
binary evolution time scale  (e.g. B{\"u}ning \& Ritter
2004). However, donor radii are hardly affected by these
short-time-scale effects and thus faithfully trace the long-term
evolution of $\dot{M}_2$

Our final mass-radius relation for all three types of CV donors is given by:   
\begin{equation}
\tiny
\frac{R_2}{R_{\odot}} = 
\left\{ \begin{array}{ll}
(0.110 \pm 0.005)\times \left(\frac{M_2}{M_{bounce}}\right)^{0.21^{+0.05}_{-0.10}} & {\rm Period~Bouncers} \\  
& \\
(0.230 \pm 0.008)\times \left(\frac{M_2}{M_{conv}}\right)^{0.64\pm0.02}  & {\rm Short-Period~CVs} \\
& \\
(0.299 \pm 0.010)\times \left(\frac{M_2}{M_{conv}}\right)^{0.67\pm0.04}  & {\rm Long-Period~CVs} \\
\end{array} \right.
\label{eq:m2r2}
\end{equation}
where the three regimes are formally defined as (i) period bouncers: 
$M_2 < M_{bounce}$; (ii) short-period CVs: $M_{bounce} < M_2 <
M_{conv}$ and $P_{orb} < P_{gap,-}$; (iii) long-period CVs: $M_{conv}
< M_2 < M_{evol}$ and $P_{gap,+} < P_{orb} < P_{evol}$. For reference,
let us also summarize the various quantities that have been assumed
\begin{equation}
\begin{array}{lll}
M_{bounce} &=& 0.063 \pm 0.009 ~ M_{\odot}\\
M_{conv} &=& 0.20 \pm 0.02 ~ M_{\odot}\\
M_{evol} &\simeq& 0.6-0.7 ~ M_{\odot} \\
P_{min} &=& 76.2 \pm 1.0 ~ {\rm min} \\
P_{gap,-} &=& 2.15 \pm 0.03 ~ {\rm hr} \\
P_{gap,+} &=& 3.18 \pm 0.04 ~ {\rm hr} \\
P_{evol} &\simeq& 5-6 ~ {\rm hr}.
\end{array}
\label{eq:defs}
\end{equation}

We note in passing that the best-fitting mass-radius relation in the
short-period regime is slightly steeper than might be suggested by
visual inspection of the data in Figure~3. This is because $M_2$ and
$R_2$ are correlated. More specifically, the appearance of the power
law index $b$ in Equation~\ref{sigma_stat} acts to push the fit away
from indeces $b \simeq \frac{1}{3}$ and thus in this case to larger
values. This can be verified by artificially increasing the
assumed intrinsic dispersion, since this forces the fit towards a
standard least-squares solution. As expected, the power law index then
tends to a slightly shallower value of 0.60 in the  short-period
regime, and the donor mass at period minimum becomes $M_{bounce} =
0.053 M_{\odot}$.

P05 did not derive a fit to the period bouncers, due to the sparseness
of the donor data in the sub-stellar regime. We completely agree with
P05 that the fit is very poorly constrained for this group: only 5
data points fall into this class in our sample, and only one of these
is clearly inconsistent with the $M_2-R_2$ relation for ``ordinary''
short-period systems. The $M_2-R_2$ relation in this regime is thus
strongly affected by our estimate for $P_{min}$ and by the parameters
inferred for short-period systems. These constraints ultimately decide
which points are included in the period bouncer class and also
uniquely fix $R_2$ at $M_{bounce}$. We have nevertheless provided
the best-fit parameters for this group, partly because we have no
reason to distrust these constraints, and partly because we want our
semi-empirical donor sequence to extend into the sub-stellar (period
bounce) regime; we would therefore rather have an uncertain
mass-radius relation than none.

We also derive some confidence in our fit to the period bouncers from
the fact that the best-fit mass-radius exponent 
in this regime turns out to be less than 1/3 (at somewhat better than
2-$\sigma$). This is a critical value, because the effective
mass-radius index of the donor determines the sign of the period
derivative of a CV. This can be seen explicitly by combining
Paczyniski's (1971) approximation for the radius of Roche-lobe-filling 
secondaries
\begin{equation}
\frac{R_2}{a} = 0.462\left(\frac{q}{1+q}\right)^{1/3}
\end{equation}
(which is valid for $q \ltappeq 0.8$) with Kepler's third law to
obtain 
\begin{equation}
\frac{\dot{P_{orb}}}{P_{orb}} = \frac{3 \xi - 1}{2} \frac{\dot{M}_2}{M_2}.
\label{pdot}
\end{equation}
Here, $\dot{M}_2 < 0$ is the mass transfer rate from the donor, and
$\xi$ is the effective mass-radius index of the donor along its
evolutionary track. If the mass-radius relation along the track can be 
described as a power law, $\xi$ is simply the power-law exponent. 
Equation~\ref{pdot} shows that the orbital period
decreases for $\xi > 1/3$, but increases for $\xi < 1/3$. With $\xi =
0.21^{+0.05}_{-0.10}$ for our period bouncers, these systems appear
to be evolving back towards larger periods -- as they should, since 
$P_{min}$ is supposed to be a {\em minimum} period along our donor
track.

\section{The Late Spectral Types of CV Donors}

We now turn our attention to the spectral types of CV donors, and
specifically to the $SpT$-$P_{orb}$ relationship. Ultimately, our goal
will be to test if donor models based on the $M_2-R_2$ relations we
have just derived are capable of reproducing the empirical
$SpT$-$P_{orb}$ relation. However, our immediate task in this section is 
mainly the construction of an updated compilation of spectroscopic $SpT$
determinations for CV donors. We will also carry out a direct
comparison between this sample and a sample of isolated MS stars. This
will allow us to establish the extent of any $SpT$ discrepancy between
CV donors and MS stars as a function of orbital period.

\subsection{An Updated Sample of Spectral Types for CV Secondaries}

B98 presented a compilation of 54 spectroscopically established donor
$SpT$s for CVs with $P_{orb} < 12$~hrs. Since then, many improved and
new $SpT$ estimates have become available. We therefore felt it was worth
repeating their analysis in order to increase the reliability and size
of the donor $SpT$ sample. Briefly, we extracted a list of all $SpT$
estimates for CVs with $P_{orb} < 12$~hrs from the latest version of
the Ritter \& Kolb  
(2003) catalog (Edition 7.6). We inspected all of the original
references for these $SpT$s and rejected all purely photometric $SpT$
determinations, and also a few spectroscopic estimates that we
considered to be less compelling. In some cases, we also carried out
additional literature searches and adjusted the $SpT$s and their
errors to provide best-bet estimates based on all of the available
evidence. Where no new information was available, we generally
retained B98's $SpT$ estimates, with two noteworthy exceptions: we were
unable to find any reliable spectroscopic $SpT$ determinations for OY
Car and QZ Aur in the literature. We therefore removed these objects
from our database. Our final compilation is provided in
Table~\ref{tab:spt} and contains 91 $SpT$s.

\begin{table*}
\begin{minipage}{140mm}
\caption{Spectral types of CV donors. This table is an update of
the compilation provided by B98. It is based, in the first instance, on
the $SpT$ estimates contained in the latest version (Edition 7.6) of the
Ritter \& Kolb (2003) catalog. However, all original source material
was inspected, and only spectroscopically determined $SpT$s that were
deemed reliable were accepted. In some cases (e.g. where more
than one $SpT$ estimate was available in the literature), the adopted
$SpT$s and/or errors were adjusted to provide best-bet estimates
based on all of the available evidence.} 
\begin{tabular}{@{}lrlclrl}
\hline
System &
$P_{orb}$ (hrs) &
Spectral Type &
\hspace*{3cm} &
System &
$P_{orb}$ (hrs) &
Spectral Type\\
\hline
RX J1951   &       11.808    &  M0   $\pm$  0.5  &  &   GY Cnc     &        4.211    &  M3   $\pm$  1    \\
UY Pup     &       11.502    &  K4   $\pm$  2    &  &   SDS J2048  &        4.200    &  M3   $\pm$  1    \\
V442 Cen   &       11.040    &  G6   $\pm$  2    &  &   V1043 Cen  &        4.190    &  M2.5 $\pm$  0.5  \\
DX And     &       10.572    &  K0   $\pm$  1    &  &   SDSS J0924 &        4.056    &  M3.5 $\pm$  1    \\
AE Aqr     &        9.880    &  K4   $\pm$  1    &  &   DO Dra     &        3.969    &  M4.25$\pm$  0.7  \\
1RXS J1548 &        9.864    &  K2   $\pm$  2    &  &   UU Aql     &        3.925    &  M4   $\pm$  1    \\
AT Ara     &        9.012    &  K2   $\pm$  0.5  &  &   CN Ori     &        3.917    &  M4   $\pm$  1    \\
RU Peg     &        8.990    &  K2.5 $\pm$  0.5  &  &   KT Per     &        3.905    &  M3.3 $\pm$  1    \\
GY Hya     &        8.336    &  K4.5 $\pm$  0.5  &  &   CY Lyr     &        3.818    &  M3.25$\pm$  1.25 \\
CH UMa     &        8.236    &  K6.5 $\pm$  1.5  &  &   VY For     &        3.806    &  M4.5 $\pm$  1    \\
MU Cen     &        8.208    &  K4   $\pm$  1    &  &   IP Peg     &        3.797    &  M4   $\pm$  0.5  \\
BT Mon     &        8.012    &  G8   $\pm$  2    &  &   QQ Vul     &        3.708    &  M4   $\pm$  0.5  \\
V1309 Ori  &        7.983    &  M0.5 $\pm$  0.5  &  &   WY Sge     &        3.687    &  M4   $\pm$  1    \\
V392 Hya   &        7.799    &  K5.5 $\pm$  0.5  &  &   RX J0944   &        3.581    &  M2   $\pm$  1    \\
AF Cam     &        7.776    &  K5.5 $\pm$  2    &  &   MN Hya     &        3.390    &  M3.5 $\pm$  0.5  \\
V363 Aur   &        7.710    &  G7   $\pm$  2    &  &   V1432 Aql  &        3.366    &  M4   $\pm$  0.5  \\
RY Ser     &        7.222    &  K5   $\pm$  1    &  &   TT Ari     &        3.196    &  M3.5 $\pm$  0.5  \\
AC Cnc     &        7.211    &  K2   $\pm$  1    &  &   MV Lyr     &        3.190    &  M5   $\pm$  0.5  \\
EM Cyg     &        6.982    &  K3   $\pm$  1    &  &   SDSS J0837 &        3.180    &  M5   $\pm$  1    \\
Z Cam      &        6.956    &  K7   $\pm$  2    &  &   AM Her     &        3.094    &  M4.25$\pm$  0.5  \\
SDSS J0813 &        6.936    &  K5.5 $\pm$  1    &  &   WX LMi     &        2.782    &  M4.5 $\pm$  2    \\
V426 Oph   &        6.848    &  K5   $\pm$  1    &  &   RX J1554   &        2.531    &  M4   $\pm$  1    \\
SS Cyg     &        6.603    &  K4.5 $\pm$  0.5  &  &   SDSS J1702 &        2.402    &  M1.5 $\pm$  1.1  \\
CW 1045    &        6.511    &  K6.5 $\pm$  1.5  &  &   QS Tel     &        2.332    &  M4.5 $\pm$  0.5  \\
CM Phe     &        6.454    &  M3.5 $\pm$  1.5  &  &   UW Pic     &        2.224    &  M4.5 $\pm$  1    \\
TT Crt     &        6.440    &  K5   $\pm$  0.75 &  &   UZ For     &        2.109    &  M4.5 $\pm$  0.5  \\
BV Pup     &        6.353    &  K3   $\pm$  2    &  &   HU Aqr     &        2.084    &  M4.25$\pm$  0.7  \\
AH Her     &        6.195    &  K7   $\pm$  1    &  &   DV UMa     &        2.063    &  M4.5 $\pm$  0.5  \\
XY Ari     &        6.065    &  M0   $\pm$  0.5  &  &   QZ Ser     &        1.996    &  K4   $\pm$  2    \\
LL Lyr     &        5.978    &  M2.5 $\pm$  1.5  &  &   AR UMa     &        1.932    &  M5.5 $\pm$  0.5  \\
AH Eri     &        5.738    &  M4   $\pm$  1    &  &   ST LMi     &        1.898    &  M5.5 $\pm$  1.5  \\
V347 Pup   &        5.566    &  M0.5 $\pm$  0.5  &  &   MR Ser     &        1.891    &  M6   $\pm$  1    \\
RW Tri     &        5.565    &  M0   $\pm$  1    &  &   CU Vel     &        1.884    &  M5   $\pm$  1    \\
EZ Del     &        5.362    &  M1.5 $\pm$  0.5  &  &   V2301 Oph  &        1.883    &  M5.5 $\pm$  1    \\
CZ Ori     &        5.254    &  M2.5 $\pm$  1.5  &  &   RZ Leo     &        1.836    &  M5   $\pm$  1    \\
AR Cnc     &        5.150    &  M5   $\pm$  1    &  &   Z Cha      &        1.788    &  M5.5 $\pm$  0.5  \\
EX Dra     &        5.038    &  M1.5 $\pm$  0.5  &  &   HT Cas     &        1.768    &  M5.4 $\pm$  0.3  \\
RX And     &        5.037    &  K4.75$\pm$  2    &  &   V834 Cen   &        1.692    &  M6.5 $\pm$  1.5  \\
AT Cnc     &        4.826    &  K7.5 $\pm$  1    &  &   VV Pup     &        1.674    &  M6.5 $\pm$  1    \\
DQ Her     &        4.647    &  M3   $\pm$  0.5  &  &   EX Hya     &        1.638    &  M4   $\pm$  1    \\
AI Tri     &        4.602    &  M2.5 $\pm$  1    &  &   HS Cam     &        1.637    &  M5   $\pm$  2    \\
Leo 7      &        4.483    &  M3   $\pm$  1    &  &   BZ UMa     &        1.632    &  M5.5 $\pm$  0.5  \\
SDSS J1553 &        4.391    &  M4.5 $\pm$  1    &  &   VY Aqr     &        1.514    &  M9.5 $\pm$  1    \\
SS Aur     &        4.387    &  M3   $\pm$  1    &  &   BC UMa     &        1.512    &  M6.5 $\pm$  0.5  \\
TW Vir     &        4.384    &  M5.5 $\pm$  0.5  &  &  	EI Psc     &        1.070    &  K5   $\pm$  1    \\
U Gem      &        4.246    &  M4.25$\pm$  0.5  &  &  	           &                 &                   \\
\hline
\end{tabular}
\label{tab:spt}
\end{minipage}
\end{table*}

\subsection{A Benchmark Main Sequence Sample}

In order to assess the departure of CV donors from thermal
equilibrium, we need a comparison sample of $SpT$s for MS stars. We
use the $[M/H] \simeq 0$ sample compiled by Beuermann et 
al. (1999; hereafter B99) for this purpose. This is based on source
data from Leggett et al. (1996; B99 Table 2) and  Henry \& McCarthy
(1993; B99 Table 3). Three unresolved binaries were removed from the
sample, and one missing spectral sub-type (GD165B: $SpT$=L4) was added
from Leggett et al. (2001). In order to better cover the late-M and L
dwarf regime, we also added a few late-type, solar metallicity objects 
from Leggett et al. (2001). These again included
some unresolved binaries, but since data is so sparse in this regime
we did not reject these objects outright. Instead, we use these
binaries only for calibrating the $SpT-(I-K)$ relation (see below),
where blending of objects with similar $SpT$ should not introduce any
serious errors. Finally, we also added the Sun ($SpT$=G2) and Gl34A 
($SpT$=G3) as calibrators, in order to extend the MS sample into the
G-dwarf regime. $SpT$s and absolute magnitudes for these two objects
were taken from B98. Overall, however, we prefer the B99 MS sample to
that used by B98 since probable low-metallicity objects have been
removed from the former.

In addition to providing an empirical comparison sample for CV donors,
the MS sample is also useful for testing and calibrating the stellar
models we will use in constructing our semi-empirical donor
sequence. Of particular importance in this context is the ability of
the models to reproduce the observed I-K colours along the MS, since
we will follow B98 in using this colour to estimate $SpT$s for the
models.
\footnote{Throughout this paper, we give UBV on the Johnson system, RI
on the Cousins system, and JHK on the CIT system.}

\begin{figure}
%\begin{minipage}{126mm}
\includegraphics[width=84mm]{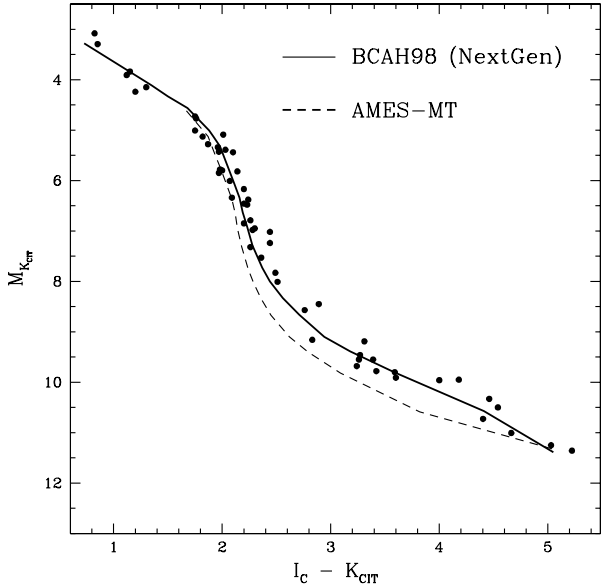}
\caption{The I-K colour-magnitude diagram for the solar-metallicity
main sequence sample. The solid line shows the 5~Gyr BCAH98 isochrone,
which is based on the NextGen atmosphere models of Hauschildt, Allard
\& Baron (1999). The dashed line corresponds to a 5~Gyr isochrone
based on a later version of these models (AMES-MT), which makes use of
an updated TiO line list (Allard, Hauschildt \& Schwenke 2000). Note
that the BCAH98 models actually provide a much better match to the data.} 
\label{fig:cmd}
%\end{minipage}
\end{figure}

So how well do up-to-date stellar models reproduce the I-K colours of
MS stars? In Figure~\ref{fig:cmd} we show the $M_K$ vs (I-K) CMD for our
MS sample, along with two sets of MS models. The first is the 5~Gyr
isochrone taken from Baraffe et al. (1998; hereafter BCAH98), with
colours calculated from the NextGen model atmospheres (Hauschildt,
Allard \& Baron 1999). 
The second, which we call AMES-MT, is an updated version
of the BCAH98 models and has been suggested to provide a somewhat better match to
observed optical colours (Allard, Hauschildt \& Schwenke 2000). The
main difference between the BCAH98 and AMES-MT models is the inclusion
of an updated TiO line list in the latter.
\footnote{Since the BCAH98 standard sequence does not provide a 
good match to solar parameters (see BCAH98), we only
use these models up to 0.8~$M_{\odot}$ and then supplement them with a
solar-calibrated $1.0~M_{\odot}$ model (taken from Table 3 in
BCAH98). As discussed by BCAH98, the solar calibration is achieved by
slightly varying the Helium abundance and mixing length
parameter. Models at lower masses are not sensitive to these parameter
changes. Note that both the AMES-MT sequence and our semi-empirical CV
donor sequence terminate at $0.7 M_{\odot}$.}

Somewhat surprisingly, Figure~\ref{fig:cmd} shows that the older BCAH98 
isochrone does a much better job of fitting the empirical $M_K$ vs
(I-K) CMD. We can only conclude that the AMES-MT models have 
achieved their improvement in the optical region at the expense of the
(near-)infrared. Given the crucial importance of the latter region for
our purposes, we use the BCAH98/NextGen models throughout this
paper. 

\begin{figure}
%\begin{minipage}{126mm}
\includegraphics[width=84mm]{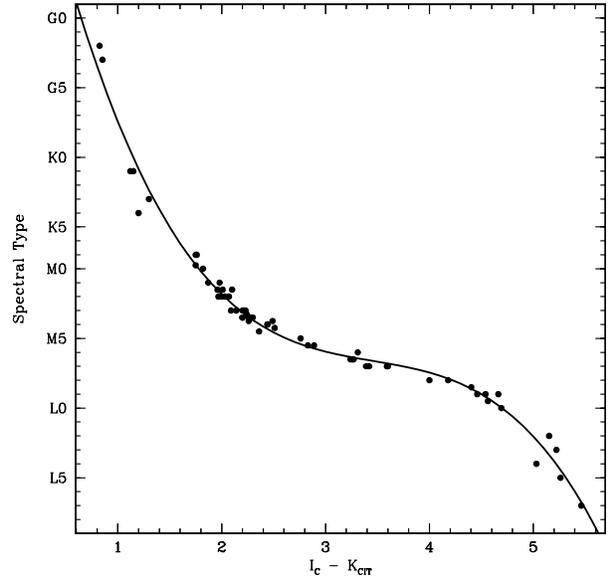}
\caption{The spectral type vs (I-K) relation for the solar metallicity
main sequence sample. The line shows the third order polynomial fit
given by Equation~\protect\ref{spt_cal}.}
\label{fig:spt_cal}
%\end{minipage}
\end{figure}

We finally revisit the $SpT$-$(I-K)$ relation defined by our MS
sample. This is shown in Figure~\ref{fig:spt_cal}. Since our sample is
substantially the same as that used by B98, their third order 
polynomial fit remains a good description of the data over most of its
range. However, since our new sample now includes a mid-L dwarf
(whereas B98's calibrating sample did not extend beyond M8), it makes
sense to redetermine the polynomial coefficients to ensure that the
fit also describes this very late-type regime. 
Figure~\ref{fig:spt_cal} shows our new calibration, which is given by
\begin{equation}
X = 56.92 - 35.21(I-K) + 9.810(I-K)^2 - 0.9450(I-K)^3, 
\label{spt_cal}
\end{equation}
where 
$SpT = L(10-X)$ for $X \leq 10$, 
$SpT = M(20-X)$ for $10 < X \leq 20$,  
$SpT = K(28-X)$ for $20 < X \leq 28$, and 
$SpT = G(38-X)$ for $28 < X \leq 38$.
Note that in order to make room for L-dwarfs, we have shifted the
$SpT$-$X$ relation (and hence the 
constant term of the polynomial fit) relative to that used by B98. As
expected, our fit parameters differ only slightly from those derived
by B98, but the new fit does indeed provide an improved match to the 
latest calibrators. The RMS scatter around the polynomial fit is less
than 0.5 sub-types for L0~$<$~$SpT$s~$<$~K5 but increases to 1-2 sub-types
outside this range.

\subsection{CV Donors vs Main Sequence Stars}
\label{donors_vs_MS}

\begin{figure*}
\begin{minipage}{180mm}
\includegraphics[width=180mm]{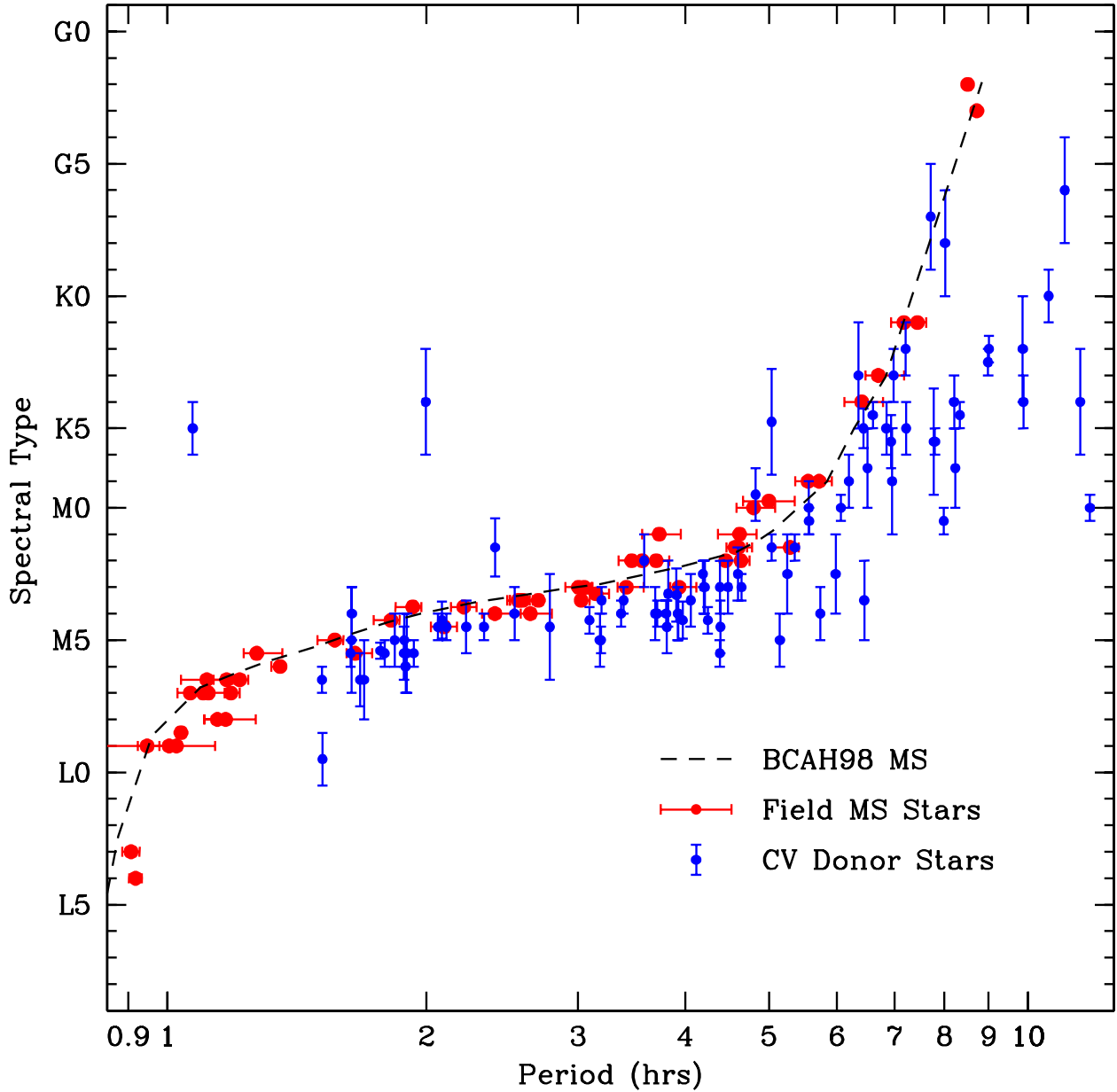}
%\begin{figure}
%\includegraphics[width=84mm]{f7.eps}
\caption{The empirical spectral type - orbital period relation for CVs
and main sequence stars. Blue points are CV donors from
Table~\ref{tab:spt}, red points are MS stars. The dashed line is the
relation predicted by the 5~Gyr BCAH98 isochrone, with the $SpT$(I-K)
calibration given by Equation~\protect\ref{spt_cal}.}
\label{fig:spt_data}
\end{minipage}
\end{figure*}

In Figure~\ref{fig:spt_data} we compare the $SpT$s of CV donors and
MS stars. We follow B98 in presenting this comparison in the
$SpT$-$P_{orb}$ plane. For this purpose, each star in the MS sample is  
assigned an orbital period on the basis of its mass and radius via the
period-density relation for Roche-lobe-filling stars; thus the
assigned $P_{orb}$ is the period of a hypothetical semi-detached
system in which the MS star is the mass donor. 

This calculation requires masses and radii for all of the stars in our
MS sample. We estimate masses from the theoretical K-band
mass-luminosity relation -- $M(M_K)$ -- predicted by the BCAH98
models. In principle, it would be preferable to use an empirical
calibration for this purpose, in order to avoid artificially forcing
the data points close to the models in the $P_{orb}-SpT$
plane. However, the best empirical $M(M_K)$ relation
(Delfosse et al. 2000) is only calibrated over the range $4.5 < M_K <
9.5$, which excludes the earliest and latest stars in our
sample. Moreover, Figure~3 in Delfosse et al. (2000) shows that
the theoretical and empirical relations agree very closely in the
well-calibrated regime, but there are hints that beyond this range,
their 5th-order polynomial fit becomes less reliable than the
theoretical models. In any case, we have checked that adopting
the Delfosse et al. $M(M_K)$ relation (even beyond its region of
validity) would not change any of our conclusions.

Several radius estimates for the stars in our MS sample are available
from B99 and Leggett et al. (2001). The bulk of our sample comes from
B99, who provide three radius estimates for each star. These are based
on (i) the $S_K(V-K)$ relation for MS stars, where $S_K$ is the K-band 
surface brightness (see also Section~\ref{sec:irmags}); (ii) the
$S_K(I-K)$ relation; (iii) the $R_2(M_K)$ relation. B99 show that all
of these estimates are probably good to better than 5\%. We adopt the
average of these estimates and use their range as a measure of the
associated error on the radius (and hence on the assigned orbital
period).

Figure~\ref{fig:spt_data} also shows the $SpT$-$P_{orb}$ relation
predicted by the BCAH98 5~Gyr isochrone. For this purpose, periods
were assigned to the models in the same way as for the MS stars
(i.e. based on the model masses and radii), and $SpT$s were estimated
from Equation~\ref{spt_cal}.

%\end{figure}

Several important results emerge
immediately. First, the $SpT$-$P_{orb}$ relation of MS
stars is fairly well described by the BCAH98 model sequence. Second,
the $SpT$s of CV donors are systematically later than those of isolated
MS stars. Importantly, this is the case across the entire $P_{orb}$
range, not just for long-period systems with $P_{orb} \gtappeq
3$~hrs, as suggested by B98. (With hindsight, we think that a
discrepancy between the $SpT$s of CV donors and MS models at short
periods was already somewhat noticeable from Figs. 4 and 5 of B98.)
Third, apart from a few outliers, the CV donors with $P_{orb} \ltappeq
5-6$~hrs define a remarkably consistent $SpT$-$P_{orb}$ sequence. This
is again evidence that most CVs do indeed follow a standard
evolutionary track with relatively little scatter. Fourth, at longer
periods, $P_{orb} \gtappeq 5-6$~hrs, the $SpT$ scatter increases
markedly, although the donors remain cooler than MS stars at the same
period.

B98 already provided a promising explanation for the appearance of the 
$SpT$-$P_{orb}$ diagram at $P_{orb} \gtappeq 5-6$~hrs. More
specifically, they showed that donors that are already somewhat
nuclear-evolved at the start of mass transfer move through just the
region of the $SpT$-$P_{orb}$ diagram that is occupied by observed
systems. The increased scatter in $SpT$s at the longest periods could
then be understood as reflecting the range of central Hydrogen
abundances in these donors at the start of mass transfer. The
viability of this idea was confirmed by Baraffe
\& Kolb (2000) and Podsiadlowski, Han \& Rappaport (2003). In the
latter work, the authors also found that the dominance of evolved
systems above $P_{orb} \simeq 5$~hrs (and the dominance of unevolved
systems at shorter periods) is actually expected, even if evolved
systems comprise only a small percentage of the total CV
population. For the record, the semi-empirical donor sequence
constructed below is intended to describe the unique evolution track
followed by {\em unevolved} donors, and we therefore limit it to
$P_{orb} \ltappeq 6$~hrs. 

We finally comment briefly on the three most striking outliers in
Figure~\ref{fig:spt_data}, namely the three systems that lie
significantly {\em above} the standard MS track with $P_{orb} <
3$~hrs. In order of
decreasing period, these systems (and their spectral type references)
are SDSS J1702+3229 (Szkody et al. 2004), QZ Ser (Thorstensen et al. 2002a)
and EI Psc (Mennickent et al. 2004; see also Thorstensen et
al. 2002b). This part of the $SpT$-$P_{orb}$ diagram should only be
populated by low-metallicity systems (B98) or again by systems in
which the donor was already significantly evolved at the onset of
mass-transfer. At least for QZ Ser and EI
Psc, the latter is the 
preferred explanation and can also account for the fact that the
orbital period of EI Psc ($P_{orb} = 64$~min) is well below the
minimum period for ``normal'' CVs (Thorstensen et al. 2002ab). SDSS
J1702+3229 is located within the period gap 
($P_{orb} = 2.4$~hrs), and the $SpT$ estimate for it is based on the
TiO-band strength measured by Szkody et al. (2004). This system is
only about 2~$\sigma$ above the ``standard'' CV donor sequence, so
additional observations will be needed to confirm if this system is
genuinely abnormal. 

\section{A Complete, Semi-Empirical Donor Sequence for CVs}

\subsection{Constructing the Donor Sequence}

We will now use our empirical $M_2-R_2$ and $SpT$-$P_{orb}$ relations 
to construct and validate a complete, semi-empirical donor sequence
for CVs. More specifically, the idea is to construct the sequence by
combining the $M_2-R_2$ relation with MS isochrones and atmosphere
models and to validate it by checking its ability to match the
observed $SpT$-$P_{orb}$ relation.

The final ingredient that is needed to carry out this program is a
relation between $M_2$ and $T_{eff}$ (and hence $SpT$) along the donor
sequence. An obvious first guess would be to assume that CV donors
continue to follow the mass-luminosity relationship defined by MS
stars (c.f. Patterson et al. 2003). However, this turns out to be incorrect. 
Instead, Kolb, King \& Baraffe (2001; see also Stehle, Ritter \& Kolb
1996; Baraffe \& Kolb 2000; Kolb \& Baraffe 2000) show that unevolved,
solar-metallicity donors should 
be expected to have the same effective temperature as MS stars of
identical mass, with essentially no dependence on the mass-loss
rate. \footnote{This statement is valid down to at least $M_2 = 0.1
M_{\odot}$. Like most aspects of the donor physics, it becomes
less robust as we approach the brown dwarf regime (Baraffe \& Kolb 
2000).}

It is now straightforward to construct our semi-empirical donor
sequence. Starting from the empirical $M_2-R_2$ relation, we can 
use the period-density relation for CV donors to obtain $P_{orb}$ for
any given donor mass. The corresponding effective temperature can be found
by interpolating on the mass-$T_{eff}$ relation in a standard MS
isochrone. The surface gravity is of course also known (from $M_2$ and
$R_2$), so absolute magnitudes in any photometric band can be obtained
by interpolating on $T_{eff}$ and $\log{g}$ in a grid of model
atmospheres and scaling to $R_2$. Finally, the $SpT$ of each
donor can be estimated from the $SpT(I-K)$ calibration
(Equation~\ref{spt_cal}).

In practice, we use the BCAH98 5~Gyr
isochrone and the corresponding NextGen model atmosphere grid down to
$T_{eff} \simeq 2000$~K. At even lower temperatures (in the brown
dwarf regime), the treatment of dust in the atmospheres becomes
important. Between $1500$~K$ \ltappeq T_{eff} \ltappeq 2000$~K we use
the 1~Gyr ``DUSTY'' isochrone of Baraffe et al. (2002; see also
Chabrier et al. 2000) and the
``AMES-DUSTY'' atmosphere models of Allard et al. (2001). At even
lower temperatures, we use the 
1~Gyr COND isochrone of Baraffe et al. (2003) and the AMES-COND  
atmosphere models of Allard et al. (2001). Physically, these two sets
of models differ 
in that dust is assumed to be present in DUSTY atmospheres, but is
assumed to have condensed out and settled in COND models. The nature
and location of the switch from DUSTY to COND conditions is poorly
understood even in isolated brown dwarfs, and represents a significant
element of uncertainty for our donor models. We use younger models to
represent brown dwarf donors since nuclear processes only stop in 
CV secondaries once they have reached $M_2 \simeq 0.07
M_{\odot}$. They are therefore effectively ``born'' as brown dwarfs at
this point. The particular choice of 1~Gyr isochrones to represent 
our brown dwarf donors is based on a comparison of the $M_2-T_{eff}$
relation to the CV donor models of Kolb \& Baraffe (1999). Isochrones
at 1~Gyr do a reasonable job at reproducing this relation down to the
lowest masses. 

We start our semi-empirical donor sequence at $M_2 < 0.70 M_{\odot}$
(or equivalently $P_{orb} \ltappeq  6.0$~hrs). As explained in 
Section~\ref{donors_vs_MS}, the longer-period CV population is likely to be
dominated by systems with evolved donors, and the evolution tracks of
such systems depend on the central Hydrogen abundance at the onset of
mass-transfer. By contrast, the donor sequence constructed here is
meant to describe the unique evolution track appropriate to unevolved
secondaries, as are likely found in the vast majority of CVs. This
particular starting location is also convenient since at this point
the donor mass-radius relation just crosses the MS
(Figure~\ref{fig:m2r2}).\footnote{This cross-over is actually the correct 
behaviour for unevolved donors, since more massive stars are 
predominantly radiative and thus {\em contract} in response to
mass-loss.}

\begin{figure*}
\begin{minipage}{160mm}
\includegraphics[width=160mm]{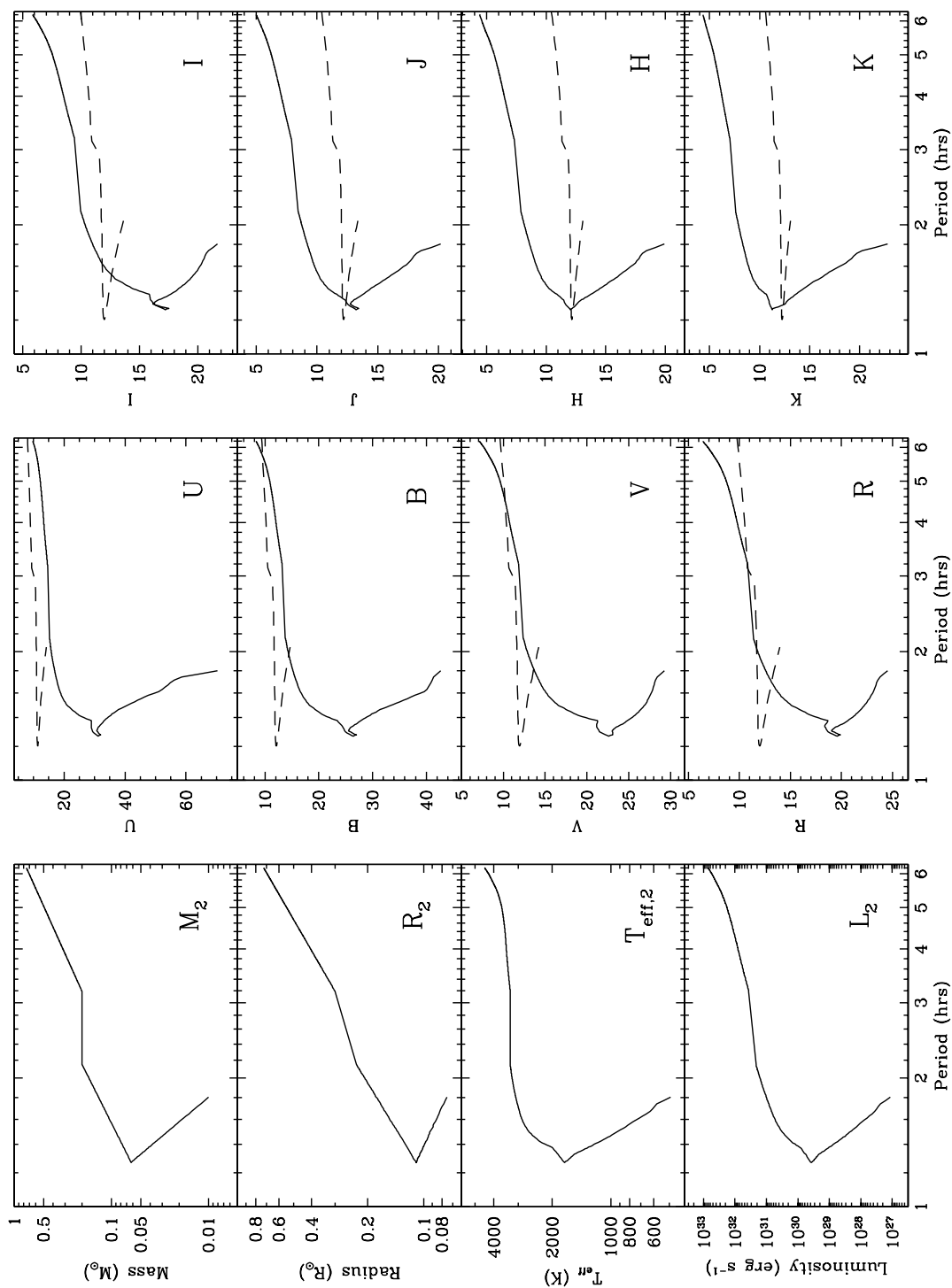}
\caption{\scriptsize Physical and photometric parameters along the CV donor
sequence. In the left column, we show the physical donor parameters
($M_2$, $T_{eff,2}$, $R_2$ and $L_2$) as a function of orbital
period. In the middle column, the solid lines show the optical absolute
magnitudes  ($M_U$, $M_B$, $M_V$ and $M_R$) as a function of
$P_{orb}$. Finally, in the right column, the solid lines correspond to
the red optical and near-infrared absolute magnitudes ($M_I$, $M_J$,
$M_H$ and $M_K$). The wiggles in the photometric parameters near
period minimum are due to switches in the model atmosphere grids used
(BCAH98, DUSTY, COND). In the middle and right
columns, we also showestimates of the absolute magnitude sequences
expected for the accretion-heated white dwarfs in CVs (dashed lines);
see text for details.}  
\label{fig:allpars}
\end{minipage}
\end{figure*}

The complete donor sequence constructed in this way, spanning the
range $0.01M_{\odot} \leq M_2 \leq 0.70 M_{\odot}$ is listed in
Table~\ref{tab:donorseq} and plotted as a function of $P_{orb}$ in
Figure~\ref{fig:allpars}. More specifically, the figure shows the 
evolution of the donor's physical parameters and also of its absolute
optical and near-infrared magnitudes. For reference, we also show in
Figure~\ref{fig:allpars} the expected 
absolute magnitude of the accretion-heated white dwarf as a function of
$P_{orb}$. The effective temperature of the WD has been calculated
following the prescription of Townsley \& Bildsten (2003), assuming a
standard theoretical CV evolution track (Rappaport, Verbunt \& Joss
1983) to yield $\dot{M}-P_{orb}$. The corresponding absolute
magnitudes were obtained by interpolating on the WD models of Bergeron
et al. (1995). These WD tracks provide a robust lower limit on 
the accretion light, since in reality the accretion disk is likely to
dominate over the WD in the optical and infrared regions.

\begin{table*}
\begin{minipage}{175mm}
\caption{The semi-empirical donor sequence for CVs. Orbital periods
are in hours, masses, radii and luminosities in solar units, and effective
temperatures in Kelvin. UBVRI magnitudes
are given on the Johnson-Cousins system (Bessell 1990), JHK magnitudes
are on the CIT system (Elias et al. 1982ab). The sequence provided
here is abbreviated. A more complete sequence, using steps of
$0.001M_{\odot}$ and including the far-infrared L, L$^\prime$ and M
bands is available in electronic form.}
\footnotesize
\begin{tabular}{@{}cccccccclccccccccc}
\hline
$P_{orb}$ &
\hspace{0.1cm} &
$M_2$ &
$R_2$ &
$T_{eff}$ &
$\log{g}$ &
$\log{L_2}$ &
\hspace{0.2cm} &
$SpT$ &
\hspace{0.2cm} &
$M_U$ &
$M_B$ &
$M_V$ &
$M_R$ &
$M_I$ &
$M_J$ &
$M_H$ &
$M_K$ \\
\hline
 1.462  & & 0.030  & 0.095  &  1009  & 4.964  & 28.50  & &     T & &  37.82  & 31.40  & 26.42  & 21.89  & 19.00  & 15.14  & 15.19  & 15.17    \\
 1.420  & & 0.035  & 0.098  &  1140  & 5.003  & 28.74  & &     T & &  35.32  & 29.55  & 25.76  & 21.33  & 18.37  & 14.42  & 14.51  & 14.41    \\
 1.384  & & 0.040  & 0.100  &  1271  & 5.037  & 28.96  & &     T & &  33.52  & 28.09  & 25.02  & 20.75  & 17.80  & 13.89  & 13.89  & 13.73    \\
 1.353  & & 0.045  & 0.103  &  1407  & 5.067  & 29.15  & &     T & &  32.22  & 26.92  & 24.25  & 20.09  & 17.13  & 13.38  & 13.32  & 13.20    \\
 1.326  & & 0.050  & 0.105  &  1543  & 5.094  & 29.33  & &     T & &  31.13  & 25.89  & 23.39  & 19.32  & 16.42  & 12.91  & 12.79  & 12.72    \\
 1.302  & & 0.055  & 0.107  &  1609  & 5.118  & 29.42  & &     T & &  30.83  & 25.64  & 23.03  & 19.09  & 16.27  & 12.78  & 12.53  & 12.41    \\
 1.281  & & 0.060  & 0.109  &  1675  & 5.141  & 29.51  & &     T & &  31.68  & 26.67  & 23.09  & 19.71  & 17.21  & 13.27  & 12.27  & 11.61    \\
 1.268  & & 0.063  & 0.110  &  1732  & 5.153  & 29.57  & &     T & &  31.22  & 26.41  & 22.63  & 19.58  & 17.21  & 13.21  & 12.04  & 11.24    \\
 1.290  & & 0.065  & 0.113  &  1784  & 5.148  & 29.65  & &  L6.7 & &  29.87  & 25.28  & 21.79  & 18.91  & 16.56  & 12.74  & 11.76  & 11.10    \\
 1.334  & & 0.070  & 0.118  &  1894  & 5.139  & 29.79  & &  L2.6 & &  28.90  & 24.42  & 21.15  & 18.36  & 15.97  & 12.32  & 11.46  & 10.90    \\
 1.377  & & 0.075  & 0.123  &  2003  & 5.131  & 29.93  & &  L3.7 & &  28.98  & 23.44  & 21.26  & 18.68  & 15.86  & 11.62  & 10.90  & 10.66    \\
 1.418  & & 0.080  & 0.128  &  2313  & 5.124  & 30.21  & &    M9 & &  24.94  & 20.92  & 18.94  & 17.02  & 14.46  & 10.93  & 10.26  &  9.96    \\
 1.457  & & 0.085  & 0.133  &  2477  & 5.116  & 30.36  & &  M7.6 & &  22.86  & 19.55  & 17.72  & 16.05  & 13.69  & 10.59  &  9.95  &  9.63    \\
 1.496  & & 0.090  & 0.138  &  2641  & 5.110  & 30.51  & &  M6.8 & &  21.03  & 18.29  & 16.59  & 15.10  & 12.96  & 10.28  &  9.66  &  9.34    \\
 1.533  & & 0.095  & 0.143  &  2726  & 5.103  & 30.59  & &  M6.5 & &  20.11  & 17.64  & 16.00  & 14.60  & 12.57  & 10.10  &  9.48  &  9.17    \\
 1.569  & & 0.100  & 0.148  &  2812  & 5.097  & 30.67  & &  M6.3 & &  19.29  & 17.04  & 15.44  & 14.11  & 12.20  &  9.93  &  9.32  &  9.00    \\
 1.638  & & 0.110  & 0.157  &  2921  & 5.086  & 30.79  & &  M5.9 & &  18.30  & 16.29  & 14.75  & 13.50  & 11.72  &  9.67  &  9.07  &  8.76    \\
 1.704  & & 0.120  & 0.166  &  2988  & 5.076  & 30.88  & &  M5.6 & &  17.70  & 15.81  & 14.30  & 13.10  & 11.40  &  9.48  &  8.88  &  8.58    \\
 1.768  & & 0.130  & 0.175  &  3055  & 5.066  & 30.96  & &  M5.3 & &  17.16  & 15.38  & 13.90  & 12.74  & 11.12  &  9.29  &  8.71  &  8.40    \\
 1.828  & & 0.140  & 0.183  &  3103  & 5.057  & 31.03  & &    M5 & &  16.76  & 15.06  & 13.59  & 12.46  & 10.89  &  9.14  &  8.55  &  8.26    \\
 1.886  & & 0.150  & 0.192  &  3151  & 5.049  & 31.10  & &  M4.8 & &  16.37  & 14.74  & 13.30  & 12.20  & 10.68  &  9.00  &  8.42  &  8.12    \\
 1.942  & & 0.160  & 0.200  &  3185  & 5.042  & 31.15  & &  M4.6 & &  16.08  & 14.50  & 13.07  & 11.99  & 10.51  &  8.87  &  8.30  &  8.00    \\
 1.996  & & 0.170  & 0.207  &  3218  & 5.035  & 31.20  & &  M4.4 & &  15.82  & 14.28  & 12.86  & 11.80  & 10.36  &  8.75  &  8.18  &  7.89    \\
 2.049  & & 0.180  & 0.215  &  3246  & 5.028  & 31.25  & &  M4.3 & &  15.60  & 14.09  & 12.68  & 11.64  & 10.22  &  8.65  &  8.07  &  7.79    \\
 2.100  & & 0.190  & 0.223  &  3269  & 5.021  & 31.29  & &  M4.2 & &  15.42  & 13.93  & 12.53  & 11.50  & 10.10  &  8.55  &  7.98  &  7.69    \\
 2.149  & & 0.200  & 0.230  &  3292  & 5.015  & 31.33  & &    M4 & &  15.24  & 13.77  & 12.38  & 11.36  &  9.98  &  8.45  &  7.88  &  7.60    \\
 3.179  & & 0.200  & 0.299  &  3292  & 4.789  & 31.56  & &  M4.2 & &  14.59  & 13.19  & 11.85  & 10.83  &  9.44  &  7.88  &  7.31  &  7.02    \\
 3.559  & & 0.250  & 0.347  &  3376  & 4.756  & 31.73  & &  M3.8 & &  13.95  & 12.58  & 11.27  & 10.29  &  8.95  &  7.47  &  6.90  &  6.63    \\
 3.902  & & 0.300  & 0.392  &  3436  & 4.729  & 31.87  & &  M3.5 & &  13.48  & 12.13  & 10.84  &  9.88  &  8.58  &  7.15  &  6.57  &  6.31    \\
 4.218  & & 0.350  & 0.434  &  3475  & 4.706  & 31.98  & &  M3.3 & &  13.13  & 11.79  & 10.51  &  9.57  &  8.29  &  6.89  &  6.31  &  6.05    \\
 4.512  & & 0.400  & 0.475  &  3522  & 4.686  & 32.08  & &  M3.1 & &  12.80  & 11.46  & 10.19  &  9.26  &  8.02  &  6.65  &  6.06  &  5.82    \\
 4.788  & & 0.450  & 0.514  &  3582  & 4.669  & 32.18  & &  M2.8 & &  12.47  & 11.13  &  9.86  &  8.96  &  7.75  &  6.42  &  5.83  &  5.60    \\
 5.050  & & 0.500  & 0.552  &  3649  & 4.653  & 32.27  & &  M2.5 & &  12.16  & 10.81  &  9.54  &  8.66  &  7.49  &  6.20  &  5.60  &  5.39    \\
 5.299  & & 0.550  & 0.588  &  3760  & 4.639  & 32.38  & &    M2 & &  11.79  & 10.40  &  9.14  &  8.28  &  7.19  &  5.97  &  5.35  &  5.16    \\
 5.537  & & 0.600  & 0.623  &  3893  & 4.626  & 32.49  & &  M1.4 & &  11.42  &  9.97  &  8.70  &  7.88  &  6.88  &  5.73  &  5.08  &  4.93    \\
 5.766  & & 0.650  & 0.658  &  4065  & 4.615  & 32.61  & &  M0.4 & &  10.96  &  9.45  &  8.19  &  7.40  &  6.54  &  5.48  &  4.83  &  4.71    \\
 5.985  & & 0.700  & 0.691  &  4239  & 4.604  & 32.73  & &    K7 & &  10.46  &  8.94  &  7.70  &  6.94  &  6.20  &  5.24  &  4.62  &  4.52    \\
 6.198  & & 0.750  & 0.724  &  4460  & 4.593  & 32.85  & &    K5 & &   9.74  &  8.34  &  7.15  &  6.43  &  5.85  &  5.00  &  4.42  &  4.35    \\
\hline
\end{tabular}
\label{tab:donorseq}
\end{minipage}
\end{table*}

One of the interesting aspects of the semi-empirical donor
sequence is the fairly sharp decline in $T_{eff}$, $L_2$, and
optical/infrared brightness in the short-period regime, but still
prior to systems reaching $P_{min}$. Beyond $P_{min}$, this trend of
course accelerates even more. In our donor track, even the 
transition point between L and T spectral types occurs before period
bounce, with spectral type L being found only in a narrow 
range of periods (1.3~hrs $\ltappeq P_{orb} \ltappeq$ 1.4~hrs). 
It should, of course, be kept in mind that our $SpT$(I-K)
calibration is relatively poorly constrained in the L/T-dwarf regime,
that the MS-based $M_2$-$T_{eff}$ calibration becomes more inaccurate
at the lowest masses, and that the 
atmosphere models become increasingly unreliable at the coolest
temperatures. For example, with our calibration, 
L-dwarfs are found at $1800$~K~ $\ltappeq ~ T_{eff} ~ \ltappeq$ ~ $2200$~K, a
somewhat narrower range than that typically quoted for isolated
L-dwarfs ($1500$~K $\ltappeq$ $T_{eff}$ $\ltappeq$~$2500$~K; Kirkpatrick
2005). On the other hand, there are also significant differences
between the effective temperatures derived for isolated L-dwarfs by 
different methods. Thus Leggett et al. (2001) found an L-dwarf
temperature range of $1800$~K$~\ltappeq T_{eff} \ltappeq 2000~$K from
spectral fitting, but $1500$~K$ \ltappeq T_{eff} \ltappeq 2200$~K based
on the estimated radii and luminosities for their sample.

Figure~\ref{fig:allpars} also shows that, in period
bouncers, even the accretion-heated white dwarf alone outshines the
donor in all optical bands and is comparable to the donor in the
infrared. All of this may explain why attempts to detect brown dwarfs 
in CVs (and especially those in suspected period bouncers) have met
with very limited success to date.

\subsection{Validating the Donor Sequence}

\begin{figure*}
\begin{minipage}{170mm}
\includegraphics[width=170mm]{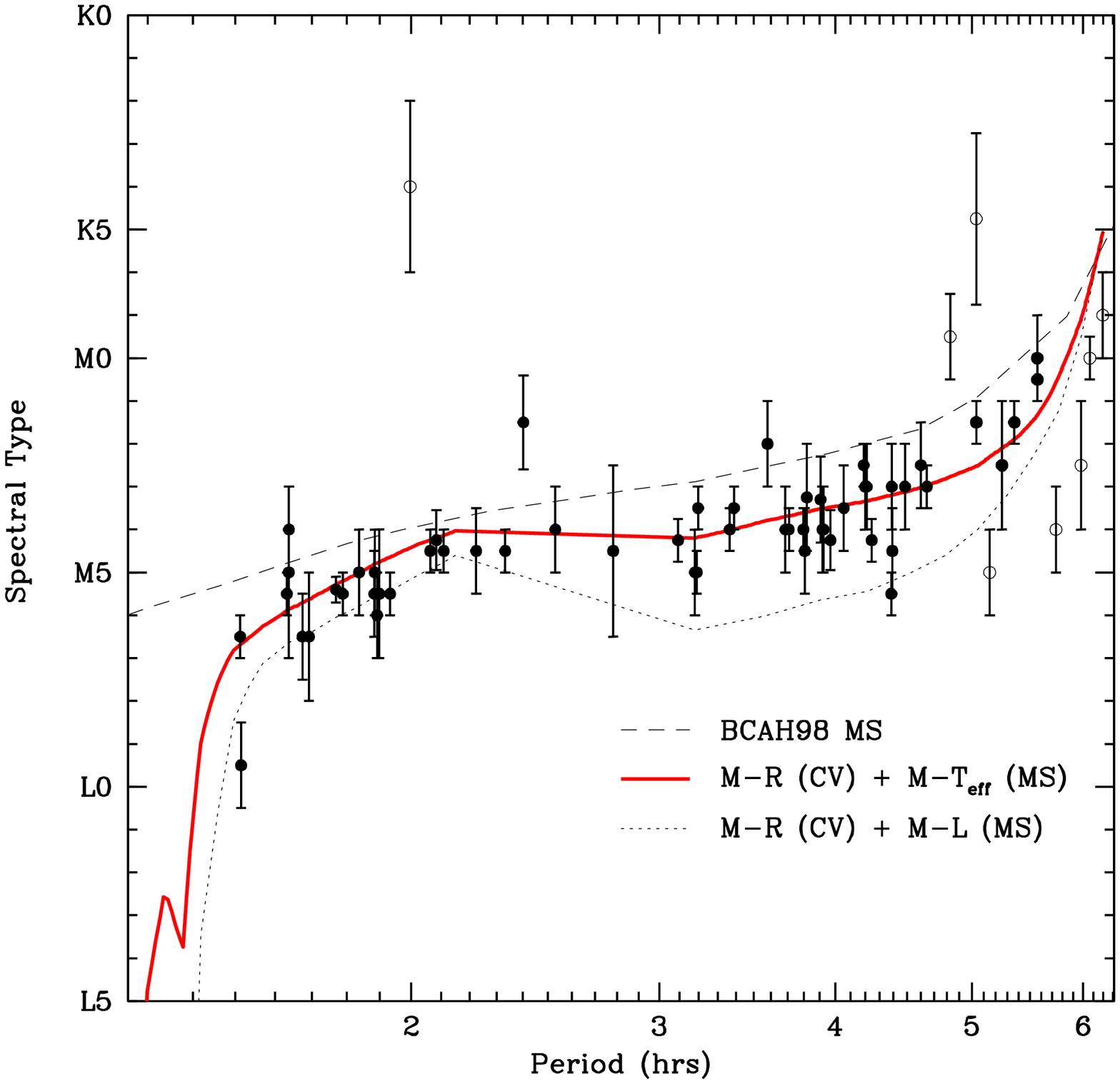}
\caption{Observed vs predicted spectral types of CV donors. Points
correspond to empirically determined $SpT$s for CVs with \protect 
$P_{orb} \protect \ltappeq  6$~hrs,
taken from Table~\protect\ref{tab:spt}. Likely evolved donors are
shown as open symbols and were excluded from the statistics quoted 
in the text. The solid red line shows the $SpT$(\protect $P_{orb}$) relation
predicted by the semi-empirical donor sequence. The dashed line shows
the relation predicted by the BCAH98 5~Gyr isochrone. The dotted line
shows the relation that would be obtained if CV donors obeyed the 
MS mass-luminosity relation, rather than the MS mass-effective
temperature relation. The kinks in the solid and dotted lines are due
to a switch between two different model atmosphere grids.}
\label{fig:spt_pretty}
\end{minipage}
\end{figure*}

So how well does our donor sequence fit the observed $SpT$-$P_{orb}$
data for CVs? The answer is provided graphically in
Figure~\ref{fig:spt_pretty}. In our view, the match to the data is
very good. The average offset between the donor sequence and
the filled data points in Figure~\ref{fig:spt_pretty} is consistent with
zero (0.1 spectral sub-types), and the RMS scatter of the data 
around the sequence is 0.9 sub-type. Most of this scatter
can be accounted for by observational uncertainties on the $SpT$
estimates: the mean (median) $SpT$ errors amongst the filled data
points is 0.86 (1) spectral sub-types. For comparison, the average
offset of the data from the standard MS is 1.2 spectral sub-types. 

It is worth stressing that the predicted sequence is not a fit to the
data, i.e. the good match between predicted and observed $SpT$s has
been achieved without any adjustable parameters. What this means is
that the empirical $M_2-R_2$ relation yields just the right amount of
radius expansion to account for the late spectral types of the
donors. Thus the results of P05 and B98 are mutually consistent. 

We also show in Figure~\ref{fig:spt_pretty} the donor sequence that would
result if we had adopted the MS mass-luminosity relation for CV donors
(rather than the correct mass-effective temperature one). As expected,
this predicts even later spectral types, since in this case the
bloated donor must reduce its $T_{eff}$ in order to maintain the MS
luminosity that would be appropriate for its mass. This sequence does
not match the data nearly as well -- the average offset from the data
is 1.4 spectral sub-types -- but it does provides a useful lower
limit on the predicted $SpT$s. 

We finally comment again on the remarkably small scatter in the
$SpT$-$P_{orb}$ diagram for unevolved CVs. In principle, we could try to
estimate the intrinsic dispersion demanded by the data in the same way
as in Section~\ref{primary}, i.e. by demanding that $\chi^2_{\nu} = 1$
for the correct $\sigma_{int,SpT}$. This would yield $\sigma_{int,SpT}
\simeq 0.4$. However, this value should perhaps not be taken at face
value, since most spectral type errors are {\em ad hoc} estimates that
cannot be expected to be Gaussian or even strictly 1-$\sigma$. A more robust
statement is that the average intrinsic dispersion must be less than
the RMS scatter of about 1 spectral sub-type. This is still quite
impressive: in the M-dwarf regime, one sub-type typically spans an
effective temperature range of only $\Delta T_{eff} \simeq
100-250$~K.

\subsection{Applying the Donor Sequence: Distances from Photometric Parallaxes}
\label{sec:irmags}

We conclude our study by exploring one obvious practical application
of the semi-empirical donor sequence, namely distance
estimation. Distances towards CVs are of fundamental importance to
virtually all areas of CV research, but are 
notoriously difficult to determine. In recent years, trigonometric
parallaxes have finally been determined for a number of systems, but
indirect methods remain the only way to obtain distance estimates for
the vast majority of CVs.  

Perhaps the most widely used indirect technique for estimating
distances towards CV is Bailey's (1981) method. This is based on the 
K-band surface brightness ($S_K$) of the secondary, defined as 
\begin{equation} 
S_K = K + 5 - 5\log{d} + 5\log{R_2/R_{\odot}}
\end{equation}
where $d$ is the distance in parsecs. $S_K$ can be calibrated against
colour (usually $(V-K)$) by using a suitable sample of MS stars. The
distance towards a CV can then be estimated if the apparent K-band
magnitude, $V-K$ colour and radius of the donor are known. In
practice, the K-band magnitude of the CV is usually assumed to be
dominated by the donor (so the resulting distance estimate 
is really a lower limit). Also, the radius is generally estimated from
the orbital period, by using the $P_{orb}$-$\rho_2$ relation
(Equation~\ref{dense}) and assuming a MS mass-radius
relation. However, the most difficult aspect of the method is the
determination of an appropriate $V-K$ value for the secondary, since
the optical flux of CVs is usually dominated by accretion light.  

Given this difficulty, much of the original promise of Bailey's method
derived from the form he inferred for the $S_K(V-K)$ relation. He
found that for M-dwarfs (and thus for almost all CV 
donors), $S_K$ was essentially constant, so that even large errors in
$V-K$ would not seriously affect the resulting distance estimates. However,
the $S_K(V-K)$ relation has since been recalibrated by Ramseyer
(1994), B99 and, most importantly, Beuermann (2000). The last of these
studies provides by far the cleanest calibration to date and shows
definitively that $S_K$ 
is {\em not} constant in the M-dwarf regime. Instead, the $S_K(V-K)$
relation is approximately linear over essentially the full range of
$SpT$s found in CVs. As a result, Bailey's original method is much less
robust than had previously been supposed, unless the contribution of
the donor to {\em both} V and K 
band fluxes can be estimated reliably from observations. This is
usually impossible, and the only way forward is then to assume
a ``typical'' donor $(V-K)$ for any given CV, based perhaps on its 
observed $SpT$ or just on its orbital period. 

Our semi-empirical donor sequence provides a way to simplify 
and improve such distance estimates. As already noted above, the 
small scatter around the $M_2-R_2$ and the $SpT-P_{orb}$ relations
implies that most CVs do, in fact, follow the unique evolution track
that is delineated by this sequence. We can therefore use the absolute
magnitudes predicted along the sequence to obtain lower limits on the
distance towards any CV, under the single assumption that the system
follows the standard track. For example, the lower limit on the
distance associated 
with a single epoch $K$-band measurement is 
\begin{equation}
\log{d_{lim}} = \frac{K - M_{K,2}(P_{orb}) + 5}{5},
\end{equation}
where $K$ is the apparent magnitude and $M_{K,2}$ is the absolute
$K$-band magnitude on the donor sequence at the CV's orbital
period. In principle, the apparent magnitude should be
extinction-corrected, but in practice this correction is usually
negligible for CVs in the infrared. If the donor contribution to the
total flux is known, the apparent magnitude of the system should of
course be replaced with that of the donor to yield an actual estimate
of the distance (rather than just a lower limit).

\begin{figure}
\includegraphics[width=84mm]{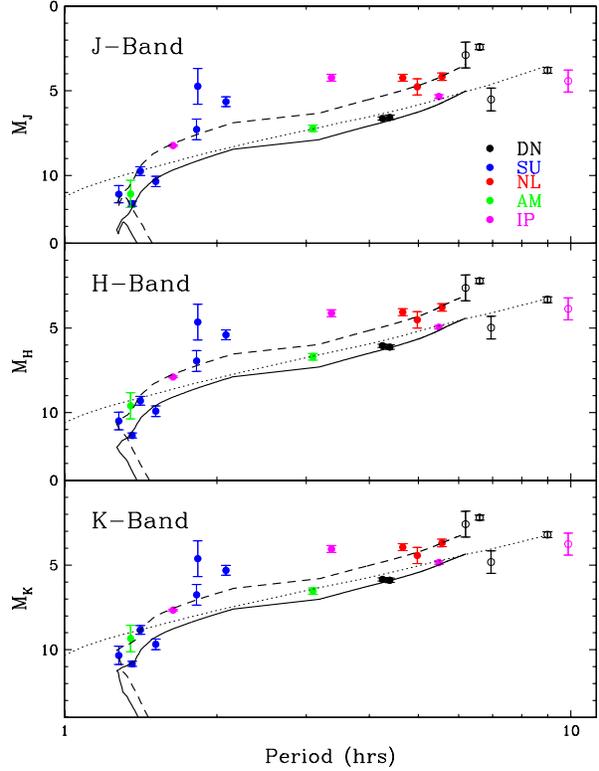}
\caption{Infrared absolute magnitudes of CVs as a function of
$P_{orb}$. Points correspond to absolute magnitudes for CVs with
trigonometric parallaxes and reliable 2MASS observations
(Table~\protect\ref{tab:ir}). Colours 
correspond to various CV sub-types, as indicated (DN = dwarf nova; SU
= SU UMa star; NL = non-magnetic nova-like; AM = AM Her star; IP =
intermediate polar). Systems with $P_{orb} > 6$~hrs are shown as open
symbols. Solid lines show the tracks predicted by the
semi-empirical donor sequence, and dotted lines show the tracks
corresponding to the BCAH98 5~Gyr isochrone. The dashed lines show the
CV donor tracks offset by the average offset between the data points
and the pure donor predictions; see text for details.}
\label{fig:irmags}
%\end{minipage}
%\end{figure*}
\end{figure}

In order to test this method, we have compiled distances and 2MASS
infrared magnitudes for all CV with trigonometric parallax
measurements. The sample is listed in Table~\ref{tab:ir} 
and contains 23 systems, 22 of which have reliable 2MASS
measurements. In Figure~\ref{fig:irmags}, the absolute 
JHK magnitudes are shown as a function of orbital period and compared
to the predictions for the donor stars from the semi-empirical
sequence in Table~\ref{tab:donorseq}. We have intentionally not
corrected the observed data points for interstellar extinction. In 
practical applications, extinction estimates will often not be
available, and our main goal here is to test how well distances may be
estimated just from single-epoch infrared magnitudes and orbital
periods. Extinction 
estimates are nevertheless listed in Table~\ref{tab:ir} and are
negligible for the sources in our sample.
\footnote{There is of course
a significant bias here, in that our trigonometric parallax sample is,
by definition, nearby and thus less heavily reddened than a more
``typical'' CV sample. However, infrared extinction is very unlikely
to ever dominate the error budget, even for more distant CVs.}

%\begin{figure*}
%\begin{minipage}{170mm}
%\includegraphics[width=170mm]{f10.eps}

Perhaps the most important point to note from Figure~\ref{fig:irmags} is
that the semi-empirical donor sequence does a nice job of tracing the
lower envelope of the data points. No CV with $P_{orb} < 6$~hrs is
significantly fainter in the infrared than the donor prediction.
%\footnote{The one system in Figure~\ref{fig:irmags} that lies somewhat
%significantly below the prediction for ``normal'', short-period
%CVs (especially in the H-band) is WZ Sge. This system is generally
%thought to be a period bouncer (e.g. Patterson 1998), which is
%consistent with its position in Figure~\ref{fig:irmags}.}
This again validates the sequence and implies that it can be used with
confidence to determine lower limits on CV distances. However, it is
also obvious from Figure~\ref{fig:irmags} that the majority of systems are
significantly brighter than the pure donor prediction, even in the
K-band. Thus in most CVs, even the infrared flux is dominated by
accretion light (cf. Dhillon et al. 2000). 

We have determined the average offsets between the donor sequence and
the data in the three infrared bands, for systems with $P_{orb} <
6$~hrs. In doing so, we assumed that none of the systems in our sample
are period bouncers, except WZ~Sge, which was removed from the sample
for the purpose of this calculation. The resulting offsets are
\begin{equation}
\begin{array}{ll}
\Delta J = 1.56  \hspace*{2cm} & \sigma_{\Delta J} = 1.25, \\
\Delta H = 1.34  \hspace*{2cm} & \sigma_{\Delta J} = 1.12, \\
\Delta K = 1.21  \hspace*{2cm} & \sigma_{\Delta J} = 1.07,
\end{array}
\end{equation}
where $\sigma_{\Delta JHK}$ is the scatter of the data around the
mean offsets. As one might expect, both the offsets and the scatter
decrease with increasing wavelength. However, even at $K$, the donor
typically contributes only 33\% of the total flux. The corresponding
contributions at J and H are 24\% and 29\%, respectively. Thus a
distance limit obtained from a single epoch infrared measurement
with no additional information would typically underestimate the true
distance by factors of 2.05 (J), 1.86 (H) and 1.75 (K). 

If it was deemed important to convert these robust lower limits into
(much less robust) distance estimates, one might consider applying the
mean offsets to the absolute donor magnitudes. We hasten to emphasize
that this procedure has no physical basis, and indeed there is no
reason to think {\em a priori} that a single offset should yield reasonable
results across the full range of orbital periods and CV
types. Nevertheless, the effect of applying these offsets to the donor
sequence is illustrated in Figure~\ref{fig:irmags} (dashed lines). The
$\sigma_{JHK}$ values indicate that, {\em for our sample}, the 
resulting distance estimates would be correct to about a factor of 1.78
(J), 1.68 (H) and 1.64 (K).

\begin{table*}
\begin{minipage}{145mm}
\caption{Distance moduli and absolute infrared magnitudes for CVs with
trigonometric parallaxes. Apparent magnitudes were taken from the final 2MASS
catalog and converted to the CIT system using the transformations
given by Carpenter (2001). Note that HV Vir has a parallax but no reliable 2MASS
detection. Note that any asymmetric errors on the distance moduli have been
symmetrized for simplicity; see the original references for the
exact uncertainties. We also provide estimated reddening values ($E_{B-V}$) from
Verbunt (1987). For reference, the associated  extinctions are 
$A_J = 0.89\times E_{B-V}$, $A_H = 0.43\times E_{B-V}$ and $A_K =
0.35\times E_{B-V}$
(Rieke \& Lebofsky 1985). All systems that are not included in Verbunt (1987) are 
very close and are therefore listed here with reddening entries of
(0.0). The distance references indicated in the last column are 1 =
Duerbeck (1999); 2 = Harrison et al. (2004); 3 = Thorstensen (2003); 4
= McArthur et al. (1999); 5 = McArthur et al. (2001); 6 = Beuermann et
al. (2004); 7 = Beuermann et al. (2003).}
\begin{tabular}{@{}ccccccccc}
\hline
System &
$P_{orb} (hrs)$ &
Type &
$m-M$ &
$M_J$ &
$M_H$ &
$M_K$ &
$E_{B-V}$ &
Ref \\
\hline 
      AE Aqr  & 9.880    &           DQ  &   5.04   $\protect\pm$  0.65   &    4.42   $\protect\pm$  0.65   &    3.87   $\protect\pm$  0.65   &    3.76   $\protect\pm$  0.65   &    (0.00)&   1     \\
      RU Peg  & 8.990    &        DN/UG  &   7.29   $\protect\pm$  0.17   &    3.78   $\protect\pm$  0.17   &    3.33   $\protect\pm$  0.17   &    3.20   $\protect\pm$  0.17   &    0.00  &   2   \\
       Z Cam  & 6.956    &        DN/ZC  &   6.06   $\protect\pm$  0.67   &    5.51   $\protect\pm$  0.67   &    4.98   $\protect\pm$  0.67   &    4.82   $\protect\pm$  0.67   &    0.00  &   3   \\
      SS Cyg  & 6.603    &        DN/UG  &   6.13   $\protect\pm$  0.15   &    2.41   $\protect\pm$  0.15   &    2.22   $\protect\pm$  0.15   &    2.19   $\protect\pm$  0.15   &    0.04  &   2   \\
      AH Her  & 6.195    &        DN/ZC  &   9.10   $\protect\pm$  0.76   &    2.89   $\protect\pm$  0.76   &    2.63   $\protect\pm$  0.76   &    2.58   $\protect\pm$  0.76   &    0.03  &   3   \\
      RW Tri  & 5.565    &           NL  &   7.79   $\protect\pm$  0.22   &    4.16   $\protect\pm$  0.22   &    3.78   $\protect\pm$  0.22   &    3.69   $\protect\pm$  0.22   &    0.10  &   4   \\
      TV Col  & 5.486    &           IP  &   7.87   $\protect\pm$  0.10   &    5.34   $\protect\pm$  0.10   &    4.95   $\protect\pm$  0.10   &    4.85   $\protect\pm$  0.10   &    0.00  &   5   \\
   V3885 Sgr  & 4.971    &           NL  &   5.21   $\protect\pm$  0.48   &    4.76   $\protect\pm$  0.48   &    4.52   $\protect\pm$  0.48   &    4.43   $\protect\pm$  0.48   &    0.00  &   1   \\
      IX Vel  & 4.654    &           NL  &   4.91   $\protect\pm$  0.20   &    4.23   $\protect\pm$  0.20   &    4.06   $\protect\pm$  0.20   &    3.94   $\protect\pm$  0.20   &    0.00  &   1   \\
      SS Aur  & 4.387    &        DN/UG  &   6.13   $\protect\pm$  0.13   &    6.57   $\protect\pm$  0.13   &    6.13   $\protect\pm$  0.13   &    5.89   $\protect\pm$  0.13   &    0.10  &   2   \\
       U Gem  & 4.246    &        DN/UG  &   5.01   $\protect\pm$  0.09   &    6.63   $\protect\pm$  0.09   &    6.05   $\protect\pm$  0.09   &    5.84   $\protect\pm$  0.09   &    0.00  &   2   \\
   V1223 Sgr  & 3.366    &           IP  &   8.61   $\protect\pm$  0.20   &    4.23   $\protect\pm$  0.20   &    4.12   $\protect\pm$  0.20   &    4.05   $\protect\pm$  0.20   &    0.15  &   6   \\
      AM Her  & 3.094    &           AM  &   4.49   $\protect\pm$  0.19   &    7.21   $\protect\pm$  0.19   &    6.70   $\protect\pm$  0.19   &    6.54   $\protect\pm$  0.19   &    (0.00)&   3     \\
      YZ Cnc  & 2.083    &        DN/SU  &   7.54   $\protect\pm$  0.29   &    5.64   $\protect\pm$  0.29   &    5.40   $\protect\pm$  0.29   &    5.31   $\protect\pm$  0.29   &    0.00  &   2   \\
      SU UMa  & 1.832    &        DN/SU  &   7.07   $\protect\pm$  1.06   &    4.73   $\protect\pm$  1.06   &    4.65   $\protect\pm$  1.06   &    4.62   $\protect\pm$  1.06   &    0.00  &   3   \\
    V893 Sco  & 1.823    &        DN/SU  &   5.95   $\protect\pm$  0.61   &    7.28   $\protect\pm$  0.61   &    6.95   $\protect\pm$  0.61   &    6.75   $\protect\pm$  0.61   &    (0.00)&   3     \\
      EX Hya  & 1.638    &           IP  &   4.05   $\protect\pm$  0.04   &    8.23   $\protect\pm$  0.05   &    7.89   $\protect\pm$  0.05   &    7.66   $\protect\pm$  0.05   &    0.00  &   7   \\
      VY Aqr  & 1.514    &     DN/SU/WZ  &   4.93   $\protect\pm$  0.30   &   10.34   $\protect\pm$  0.30   &    9.91   $\protect\pm$  0.31   &    9.68   $\protect\pm$  0.31   &    (0.00)&   3     \\
       T Leo  & 1.412    &        DN/SU  &   5.02   $\protect\pm$  0.26   &    9.74   $\protect\pm$  0.26   &    9.30   $\protect\pm$  0.26   &    8.83   $\protect\pm$  0.26   &    (0.00)&   3     \\
      HV Vir  & 1.370    &     DN/SU/WZ  &   8.31   $\protect\pm$  1.37   &   --                  &   --                  &   --                  &    --    &   3       \\
      WZ Sge  & 1.361    &     DN/SU/WZ  &   3.19   $\protect\pm$  0.14   &   11.66   $\protect\pm$  0.15   &   11.35   $\protect\pm$  0.15   &   10.83   $\protect\pm$  0.15   &    0.00  &   2   \\
      EF Eri  & 1.350    &           AM  &   6.06   $\protect\pm$  0.77   &   11.09   $\protect\pm$  0.80   &    9.60   $\protect\pm$  0.78   &    9.33   $\protect\pm$  0.79   &    (0.00)&   3     \\
      GW Lib  & 1.280    &     DN/SU/WZ  &   5.09   $\protect\pm$  0.51   &   11.10   $\protect\pm$  0.51   &   10.49   $\protect\pm$  0.52   &   10.33   $\protect\pm$  0.54   &    (0.00)&   3     \\
\hline
\end{tabular}
\protect\label{tab:ir}
\end{minipage}
\end{table*}

\section{Discussion and Conclusions}

The main goal of this paper has been the construction of a complete,
semi-empirical donor sequence for CVs that is based on our best
understanding of donor properties and is consistent with all key 
observational constraints. This donor sequence is provided in
Table~\ref{tab:donorseq} and is based primarily
on an updated version of the mass-radius relationship for CV 
secondaries determined by P05. It also relies on a MS-based
mass-effective temperature relation that has been shown 
to be appropriate for CV donors by theoretical work, and on up-to-date
stellar atmosphere models. By design, the donor track also reproduces 
the observed locations of the period gap and the period minimum, and
extends beyond this to the period bouncer
regime. We have shown that this sequence provides an excellent match
to the observed $SpT$-$P_{orb}$ relation for unevolved CV donors
(i.e. $P_{orb} \ltappeq 6$~hrs), and that it correctly traces the lower
envelope of the $M_{JHK}-P_{orb}$ distributions of CVs.

Along the way, we have revisited and updated two important
results on CV donors by P05 and B98. Regarding the former study, we
have carried out a full, independent analysis of the superhumper and
eclipser data that was used by P05 to construct their empirical
$M_2-R_2$ relation. We constructed and used our own $q-\epsilon$ 
calibration for superhumpers, verified the validity of the constant
$M_1$ assumption on which the method rests, imposed the locations of
the period gap and the period minimum as constraints on the
mass-radius relation and used a self-consistent fitting technique to
extract optimal parameters for the $M_2-R_2$ relations for
long-period CVs, short-period CVs and period-bouncers. 

Regarding the study by B98, we have updated their $SpT$ data base for
CV secondaries. Our new sample contains 91 CV 
donors with spectroscopically-determined $SpT$ estimates (compared to 54
donors in the B98 sample). Comparing the $SpT$-$P_{orb}$ distribution to
that of MS stars, we find that CV donors have later spectral types at
all orbital periods. This extends the finding of B98 to the
short-period regime ($P_{orb} \ltappeq 3$~hrs). 

We find that there is remarkably little intrinsic scatter about
both the mean $M_2-R_2$ and 
$SpT-P_{orb}$ relations for CVs with $P_{orb} < 6$~hrs: probably no
more than a few percent in $R_2$ and less than 1 spectral
sub-type. This suggests that, on long time-scales, most CVs do indeed
follow a unique evolution track, just as is theoretically expected
(e.g. Kolb 1993; Stehle, Ritter \& Kolb 1996). The large scatter in
luminosity-based accretion  
rate estimates at fixed $P_{orb}$ (e.g. Patterson 1984) is therefore
probably caused by fluctuations around the mean mass transfer rate on
time-scales that are short compared to the time-scale on which the donor
loses mass and the binary evolves (see B{\"u}ning \& Ritter 2004 and
references therein). In principle, the
empirical $M_2-R_2$ relationship for CV donors can be inverted to
infer the long-term average mass-transfer rate as a function of
orbital period; this will be the subject of a separate paper. 

An important feature of our donor sequence is a sharp decline in
effective temperature, luminosity and optical/IR brightness, well
before the period minimum is reached. In fact, the L/T transition 
essentially coincides with $P_{min}$ on the evolution track (although
this statement depends somewhat on the validity of our spectral type
calibration in a sparsely populated regime). Beyond $P_{min}$,
even the accretion-heated WD alone is expected to outshine the donor
at optical wavelengths and it is comparably bright to the donor even
in the infrared. All of this helps to explain why brown dwarf donors have
proven so difficult to detect in CVs, especially in suspected period
bouncers. 

Finally, we have looked at one obvious application of our donor
sequence, the determination of distance estimates from infrared
photometry. We have shown that robust lower limits can indeed 
be obtained, but that, in our calibration sample, these are typically
a factor of two smaller than the true distances. Thus, even in the
infrared, the donor typically contributes only $\simeq$30\% of the total
flux. Distance estimates that explicitly allow for this average donor
contribution are typically good to about a factor of 2 in our
calibration sample. 

\section*{Acknowledgments} 
I am extremely grateful to Joe Patterson and Isabelle
Baraffe for their input and many useful discussions. Isabelle Baraffe
also kindly provided some of the isochrones that are used in this
paper. I am also grateful to the referee, Robert Connon Smith, for his 
careful and useful report. Finally, thanks to France Allard for making
NextGen, AMES-DUSTY and AMES-COND model atmosphere grids available via
her website.

\appendix

\section{Calibrating the \MakeLowercase{$\epsilon-q$} Relation}

Here we provide some additional details on the way in which we
calibrated the $\epsilon-q$ relation. As already noted in 
Section~\ref{sec:calib}, we rely on the same calibrators as P05 (given in
their Table~7), but there are a few differences between their
treatment of the problem and ours. First, 
the two calibrators with the most precise mass ratios are XZ
Eri and DV UMa (Feline et al. 2004). In both cases, Feline et
al. obtained their final $q$-estimates as weighted averages over the
mass ratios determined in 3 different photometric bands. However, we
found the individual $q$-estimates for both systems were not quite
consistent at the level expected based on their internal errors. We
therefore re-averaged the individual $q$ values, but allowed
explicitly for just enough intrinsic dispersion to
make the values in all bands consistent with each other. The resulting
new mass ratios were $q = 0.111 \pm 0.03$ (XZ Eri) and $q = 0.152 \pm
0.003$ (DV UMa). The implied changes to the mass ratios are relatively
small, but the increase in the formal errors is important (since these
two data points would otherwise be given too much weight in the
calibration). Second, we took note of the approximate upper limit on
the mass ratio of BB Dor suggested by P05 ($q < 0.38$), but did not
explicitly enforce it in our fits. Instead, we decided to check {\em a
posteriori} if our fits significantly violated this approximate
constraint or not. Third, we chose to treat $q$ (rather than
$\epsilon$) as the independent variable of the problem, since the
ultimate goal is to estimate $q$ from a measured $\epsilon$. Fourth,
we experimented with a variety of different families of 2-parameter 
curves, including linear fits (with constant term), quadratic fits
(with no constant term) and power laws.

We found that the preferred functional form for the $\epsilon-q$
relation is linear, i.e. $q = a + b \epsilon$, with $a$ and $b$
constants. We obtained optimal estimates of $a$ and $b$ by minimizing
the $\chi^2$ of the model with respect to the data, allowing
explicitly for errors in both $q$ and $\epsilon$ and also for
the possibility of intrinsic scatter around the relation,
$\sigma_{int}$. The goodness-of-fit metric then becomes
\begin{equation}
\chi^2 = \sum_{i=1}^N \frac{(q - a - b \epsilon)^2}{\sigma^2_{q} +
b^2 \sigma_{\epsilon}^2 + \sigma_{int}^2}.
\label{chi1}
\end{equation}
The $\epsilon-q$ relation given in the main body is the minimum
$\chi^2$ solution. It does not significantly violate the approximate
upper limit on BB Dor's mass ratio and achieves a statistically
acceptable $\chi^2_\nu = 1.03$ with $\sigma_{int} = 0$. Thus any
intrinsic scatter 
around the calibrating relation must be small compared to the
statistical errors on the data point. P05 explicitly added an extra
5\% uncertainty on both $q$ and $\epsilon$ when carrying out their
fits. We appreciate their rationale for doing this (allowing for
variability in superhump periods and 
unaccounted-for external errors in published mass ratio estimates);
indeed, we accounted for the possibility of a non-zero $\sigma_{int}$
in our fits for exactly the same reasons. However, it simply turns out
that such additional error contributions are not actually demanded by
the data. 

\section{Testing the Assumption of Constant Primary Mass}
\label{app:primary}

In order to test if there is evidence for evolution of $M_1$, we
fitted a straight line to the $N=16$ eclipser data points with $P_{orb}
< 6$~hrs given in Table~7 in P05. Orbital period errors are negligible
compared to the errors on $M_1$ and to the intrinsic dispersion, so we
used the $\chi^2$ metric
\begin{equation}
\chi^2 = \sum_{i=1}^N \frac{(M_1 - a - b P_{orb})^2}{\sigma^2_{M_1} +
  \sigma^2_{int}}.
\label{chi2}
\end{equation}
Here, $\sigma_{M_1}$ is the error on $M_1$ and $\sigma_{int}$ is the
intrinsic dispersion around the fit. The intrinsic dispersion is set
to the value needed to obtain a reduced $\chi^2_{\nu} \simeq 1$, where
$\nu = N - 2$; for the linear fit, this value is $\sigma_{int} =
0.16 M_{\odot}$. The corresponding constraints on the slope and intercept
are shown in the top panel of Figure~\ref{fig:m1}, and the best-fitting
line itself is plotted in the bottom panel. There is certainly
substantial scatter among the $M_1$ values, but the best-fitting slope
is not significantly different from zero. Thus there is no convincing
evidence for a trend of $M_1$ with $P_{orb}$. The best-fitting
constant $M_1$, using the same $\chi^2$ metric, but with slope fixed
at $b = 0$, yields a mean WD mass of $M_1 = 0.75 \pm 0.05 M_{\odot}$. It
requires the same level of intrinsic scatter as the linear fit,
i.e. $\sigma_{int} = 0.16 M_{\odot}$. 

These results may seem at odds with the studies of Webbink (1990)
and Smith \& Dhillon (1998), both of whom suggested that CVs below the gap
tend to have lower WD masses than those above the gap. More
specifically, Smith \& Dhillon [Webbink] derived weighted mean masses
of  $M_1 = 0.66 \pm 0.01 M_{\odot}$ [$0. 66 \pm 0.01 M_{\odot}$] below the gap a
nd $M_1 =
0.78 \pm 0.02 M_{\odot}$ [$0. 81 \pm 0.04 M_{\odot}$] above the gap. Two points
are important to
note here. First, the weights that were used in calculating these
weighted means were the inverse variances of the formal errors on the
individual data points. Second, the quoted uncertainties on the
means are the formal errors on the weighted means, rather than the
dispersion of the points around the means. If we calculate the same
quantities for our sample (in which there are 8 short- and 8
long-period systems), we find weighted means with formal errors of
$M_1 = 0.642 \pm 0.007 M_{\odot}$ below the gap and $M_1 = 0.72 \pm
0.02 M_{\odot}$ above
the gap. So if we had used these
measures, there would appear to be a significant difference between
short- and long-period systems for our sample, too.

The resolution of this apparent paradox is simple. As already
noted by Smith \& Dhillon, the weighted means and their formal errors
are quite misleading, since they can be totally dominated by a few
systems with small formal uncertainties. If
the true intrinsic dispersion among the sample being averaged is
larger than these formal uncertainties -- as it clearly is here -- then the
formal error on the weighted mean (and even the weighted mean itself)
is rather meaningless. From a statistical point of view, such a sample
violates the basic premise underlying the calculation of the
weighted mean, which is that each data point is drawn from a
distribution with the same mean, with a variance given solely by the
formal error.

Smith \& Dhillon also give the unweighted means and dispersions for
their sample, which are $M_1 = 0.69 \pm 0.13 M_{\odot}$ below
the gap and $M_1 = 0.80 \pm 0.22M_{\odot} $ above the gap. For our sample, the
corresponding numbers are $M_1 = 0.74 \pm 0.20 M_{\odot}$ below the gap and $M_1
= 0.78 \pm 0.16 M_{\odot}$ above the gap. This already shows that the
differences between short- and long-period systems are much less
significant than suggested by the formal weighted means and errors,
and that any real trend with $P_{orb}$ that might be present is small
compared to the intrinsic dispersion.

However, the best way to estimate the means, error on the means and intrinsic
dispersion is via the statistic given by Equation~\ref{chi2} (again
with slope fixed at $b = 0$). This accounts explicitly for {\em both} the formal
errors and intrinsic dispersion, with the latter being adjusted to
yield a reasonable $\chi^2_{\nu}$. If we do this separately for
short- and long-period systems in our sample, we find  $M_1 = 0.73 \pm
0.07 M_{\odot}$ with $\sigma_{int} = 0.19 M_{\odot}$ below the gap and $M_1 = 0.
77 \pm
0.06 M_{\odot}$ with $\sigma_{int} = 0.15 M_{\odot}$ above the gap. As expected,
 these
numbers are not significantly different from each other anymore.

\section{Fitting the $M_2-R_2$ Relation}

We wish to obtain an optimal estimate of the mass-radius relation of CV
donors from a set of ($M_2,R_2$) pairs. As noted in Section~\ref{sec:fitting},
the key challenges are to acccount for the correlated nature of the
errors on $M_2$ and $R_2$ and to self-consistently impose the external
constraints derived from the locations of the period gap and the
period minimum. In addition, we also need to allow for intrinsic
scatter around the mass-radius relation.

We begin by defining the analytical form of the mass-radius relation 
we wish to fit to the data. Based on inspection of the data in
Figure~\ref{fig:m2r2}, we will follow P05 and describe the mass-radius
relation as a power law,
\begin{equation}
\frac{R_{mod}}{R_{ref}} = \left(\frac{M}{M_{ref}}\right)^b.
\label{model}
\end{equation}
Here and below, we drop the subscript ``2'' on the mass and radius in
order to keep the notation transparent. In practice, we
carry our fits out in log-space, where the power law transforms into
the linear relation 
\begin{equation}
\log{R_{mod}} = \log{R_{ref}}  \; 
+ b \log{M} \;
- b \log{M_{ref}}.
\end{equation}

In the absence of additional constraints, both $R_{ref}$ and $b$
would be free parameters of this model. However, as discussed in 
Section~\ref{sec:constraints}, we will demand that $M = M_{conv}$ 
at $P_{orb} = P_{gap,\pm}$ and $M = M_{bounce}$ at $P_{orb} =
P_{min}$. The easiest way to accomplish this is to adopt $M_{ref} =
M_{conv}$ for long- and short-period CVs and $M_{ref} = M_{bounce}$
for period bouncers. With these choices of reference mass, $R_{ref}$ 
is fixed (to within some error) by the period-density relation for
Roche-lobe filling stars (Equation~\ref{r2_2}). Thus for long
period CVs, we have 
\begin{equation}
\begin{array}{l}
\log{R_{ref}} =  \log{0.2361 R_{\odot}} \;
+ (2/3)\log{P_{gap,+}(hr)} + \\
\hspace*{0.75cm}  (1/3)\log{M_{conv}} \; 
- (1/3)\log{M_{\odot}}.
\label{ref}
\end{array}
\end{equation}
Similar relations hold for short-period CVs and period bouncers. Note
that in all these relations, $P_{gap,\pm}$(or $P_{min}$) and
$M_{conv}$ (or $M_{bounce}$) are empirical estimates with associated
unertainties. As discussed further below, these uncertainties
translate into a systematic error on each data point that needs to be
accounted for in the final fit.

However, let us first deal with the correlation between the $M$ and
$R$ estimates for a given system. The easiest way to achieve this is
to work exclusively in terms of $M$. For example, consider a particular
data point ($M_{i}$, $R_{i}$, $P_{orb,i}$) in the short-period
regime (the corresponding derivation for long-period systems and
period bouncers is identical). The residual between this point and the
model prediction is $\Delta_i = \log{R_{i}} - \log{R_{mod,i}(M_i)}$. Using
Equations~\ref{r2_2}, ~\ref{model} and~\ref{ref}, this can be rewritten as 
\begin{equation}
\begin{array}{l}
\Delta_i = \frac{2}{3}\log{P_{orb,i}} \;
- \frac{2}{3}\log{P_{gap,-}} + \\
\hspace*{0.75cm} \left(\frac{1}{3} - b\right)\log{M_i} \;
- \left(\frac{1}{3} - b\right)\log{M_{conv}}.
\label{res}
\end{array}
\end{equation}
Let us ignore the systematic errors arising from the uncertainties on
$M_{conv}$ and $P_{gap,-}$ for the moment and consider only the
statistical variance on $\Delta_i$. This is 
\begin{equation}
\sigma_{stat,i}^2 = \left(\frac{1}{3} - b\right)^2
\label{sigma_stat}
\end{equation}
where we have neglected the error on $P_{orb,i}$ which is always much
smaller than that on $M_i$. An optimal estimate for $b$ could then be
obtained by minimizing the usual $\chi^2$ statistic
\begin{equation}
\chi^2 = \sum_{i=1}^{N} \frac{\Delta_i^2}{\sigma_{stat,i}^2}, 
\end{equation}
or, if we additionally want to allow for an intrinsic variance
$\sigma_{int}^2$, 
\begin{equation}
\chi^2 = \sum_{i=1}^{N} \frac{\Delta_i^2}{\sigma_{stat,i}^2 +
  \sigma_{int}^2}, 
\label{chistat}
\end{equation}
where $N$ is the number of data points. 

We should also face up to the uncertainties on $M_{conv}$ and
$P_{gap,-}$. It is tempting to simply account for these in the denominator
of Equation~\ref{chistat}. However, this would be incorrect, since any
change in these quantities affects all residuals in the same
way: $\sigma_{\log{M}_{conv}}$ and $\sigma_{\log{P}_{gap,-}}$ are {\em
systematic} errors in this context. Let us define the systematic
variance on each residual as
\begin{equation}
\sigma_{sys}^2 = \frac{4}{9} \sigma_{\log{P}_{gap,-}}^2 \; + \; 
\left(\frac{1}{3} - b\right)^2 \sigma_{\log{M_{conv}}}^2.
\end{equation}
It is then possible to define a new $\chi^2$ statistic that has the
correct distribution and accounts for both $\sigma_{stat,i}$ and
$\sigma_{sys}$. Following Stump et al. (2002), we find that in our
case, this statistic is 
\begin{equation}
\chi^2 = \sum_{i=1}^{N} \frac{\Delta_i^2}{\sigma_{stat,i}^2 + \sigma_{int}^2} 
\; - \; \frac{\left[\sum_{i=1}^{N} \frac{\Delta_i \sigma_{sys}}{\sigma_{stat,i}^2 + \sigma_{int}^2}\right]^2}
{1 + \sum_{i=1}^{N} \frac{\sigma_{sys}^2}{\sigma_{stat,i}^2 + \sigma_{int}^2}}.
\label{chistatsys}
\end{equation}

The only uncertainties we have not explicitly accounted for 
in our fits are (i) the error on the mean white dwarf mass adopted for
superhumpers\footnote{Note that we are specifically referring to the
error on the mean here, not the intrinsic dispersion of the data around
this mean; for details on the distinction, see Section~\ref{primary} and
Appendix~\ref{app:primary}.}
and (ii) the errors on the parameters of the $\epsilon-q$
calibration. As discussed in Sections~\ref{primary} and
\ref{sec:calib}, these uncertainties also translate into systematic errors 
on the data points. In principle, it would be possible to formally
account for these  
systematics in a similar way as we just did for the errors on
$P_{gap,\pm}$ 
and $M_{conv}$, for example. However, in practice, such a treatment
would be considerably more difficult and (in our view) a little
pointless. The added difficulty arises partly because these new
systematics do not affect all data points in a given mass/period regime
(i.e. only the superhumpers, but not the eclipsers); the error on the
calibration parameters, in particular, does not even affect 
all superhumpers in the same way. We feel the additional complexity
this would introduce is not warranted. After all, regarding (i), we
already know that the error on the mean $M_1$ is small compared to
intrinsic dispersion around this mean, which we do account for (see
Section~\ref{primary}). And regarding (ii), we already noted in
Section~\ref{sec:calib} that the systematic uncertainty associated
with the $\epsilon-q$ relation is small in the well-calibrated regime, 
but not necessarily well-described by the formal parameter
uncertainties in the poorly-calibrated regime. Rather than attempt to
include this poorly defined systematic in our fits, we thus prefer to
simply emphasize its existence.  

The optimal estimate of the power-law exponent $b$ is obtained by
minimizing Equation~\ref{chistatsys}. For long- and short-period
systems, we determine the appropriate level of $\sigma_{int}$ by
requiring that the reduced $\chi^2$ should be equal to one; for period
bouncers, the reduced $\chi^2$ is slightly less than one even without
any intrinsic dispersion. However, this is probably more of a reflection of the
sparseness of the data in this regime than of any true constraint on
$\sigma_{int}$. We therefore conservatively adopt the value of
$\sigma_{int}$ obtained for short-period CVs for the period bouncers
as well. Errors on the power law exponents can be estimated in the
usual way. For example, the 1-$\sigma$ confidence interval around $b$
corresponds to the range of exponents for which $\chi^2 \leq
\chi^2_{min} + 1$, where $\chi^2_{min}$ is the lowest value of
$\chi^2$.

In order to test this method, we have carried out Monte Carlo
simulations. Thus we created fake data sets with known slopes and
subject to all of the errors we are trying to account for in our 
fits. In these simulations, the method did a good job in recovering
the correct slope, and the distribution of the recovered slopes had a
dispersion consistent with the estimated error on the slope.

\bsp

\label{lastpage}


\begin{thebibliography}{}

\bibitem[Allard et al.(2000)]{2000ApJ...540.1005A} Allard, F., Hauschildt, 
P.~H., \& Schwenke, D.\ 2000, ApJ, 540, 1005 

\bibitem[Allard et al.(2001)]{2001ApJ...556..357A} Allard, F., Hauschildt, 
P.~H., Alexander, D.~R., Tamanai, A., \& Schweitzer, A.\ 2001, ApJ, 556, 
357 

\bibitem[Bailey(1981)]{1981MNRAS.197...31B} Bailey, J.\ 1981, MNRAS, 197, 
31 

\bibitem[Baraffe et al.(1998)]{1998A&A...337..403B} Baraffe, I., Chabrier, 
G., Allard, F., \& Hauschildt, P.~H.\ 1998, A\&A, 337, 403 

\bibitem[\protect\citeauthoryear{Baraffe \&
Kolb}{2000}]{2000MNRAS.318..354B} Baraffe I., Kolb U., 2000, MNRAS, 318, 354

\bibitem[Baraffe et al.(2002)]{2002A&A...382..563B} Baraffe, I., Chabrier, 
G., Allard, F., \& Hauschildt, P.~H.\ 2002, A\&A, 382, 563 

\bibitem[Baraffe et al.(2003)]{2003A&A...402..701B} Baraffe, I., Chabrier, 
G., Barman, T.~S., Allard, F., \& Hauschildt, P.~H.\ 2003, A\&A, 402, 701 

\bibitem[Bergeron et al.(1995)]{1995PASP..107.1047B} Bergeron, P., 
Wesemael, F., \& Beauchamp, A.\ 1995, PASP, 107, 1047 

\bibitem[Bessell(1990)]{1990PASP..102.1181B} Bessell, M.~S.\ 1990, PASP, 
102, 1181 

\bibitem[Beuermann et al.(1998)]{1998A&A...339..518B} Beuermann, K., 
Baraffe, I., Kolb, U., \& Weichhold, M.\ 1998, A\&A, 339, 518 

\bibitem[Beuermann et al.(1999)]{1999A&A...348..524B} Beuermann, K., 
Baraffe, I., \& Hauschildt, P.\ 1999, A\&A, 348, 524 

\bibitem[Beuermann(2000)]{2000NewAR..44...93B} Beuermann, K.\ 2000, New 
Astronomy Review, 44, 93 

\bibitem[Beuermann et al.(2003)]{2003A&A...412..821B} Beuermann, K., 
Harrison, T.~E., McArthur, B.~E., Benedict, G.~F., G{\"a}nsicke, B.~T.\ 
2003, A\&A, 412, 821 

\bibitem[Beuermann et al.(2004)]{2004A&A...419..291B} Beuermann, K., 
Harrison, T.~E., McArthur, B.~E., Benedict, G.~F., G{\"a}nsicke, B.~T.\ 
2004, A\&A, 419, 291 

\bibitem[B{\"u}ning \& Ritter(2004)]{2004A&A...423..281B} B{\"u}ning, A., 
\& Ritter, H.\ 2004, A\&A, 423, 281 

\bibitem[Carpenter(2001)]{2001AJ....121.2851C} Carpenter, J.~M.\ 2001, AJ, 
121, 2851 

\bibitem[Chabrier et al.(2000)]{2000ApJ...542..464C} Chabrier, G., Baraffe, 
I., Allard, F., \& Hauschildt, P.\ 2000, ApJ, 542, 464 

\bibitem[Delfosse et al.(2000)]{2000A&A...364..217D} Delfosse, X., 
Forveille, T., S{\'e}gransan, D., Beuzit, J.-L., Udry, S., Perrier, C., \& 
Mayor, M.\ 2000, A\&A, 364, 217 

\bibitem[Dhillon et al.(2000)]{2000MNRAS.314..826D} Dhillon, V.~S., 
Littlefair, S.~P., Howell, S.~B., Ciardi, D.~R., Harrop-Allin, M.~K., \& 
Marsh, T.~R.\ 2000, MNRAS, 314, 826 

\bibitem[Duerbeck(1999)]{1999IBVS.4731....1D} Duerbeck, H.~W.\ 1999,
  IBVS, 4731, 1 

\bibitem[bla]{bla} Echevarr\'{i}a J.\ 1983, Rev.Mex.Astron.Astrof. 8, 109

\bibitem[Elias et al.(1982)]{1982AJ.....87.1029E} Elias, J.~H., Frogel, 
J.~A., Matthews, K., \& Neugebauer, G.\ 1982a, AJ, 87, 1029 

\bibitem[Elias et al.(1982)]{1982AJ.....87.1893E} Elias, J.~H., Frogel, 
J.~A., Matthews, K., \& Neugebauer, G.\ 1982b, AJ, 87, 1893 

\bibitem[Feline et al.(2004)]{2004MNRAS.355....1F} Feline, W.~J., Dhillon, 
V.~S., Marsh, T.~R., \& Brinkworth, C.~S.\ 2004, MNRAS, 355, 1 

\bibitem[\protect\citeauthoryear{Friend et al.}{1990}]{1990MNRAS.246..637F} 
Friend M.~T., Martin J.~S., Connon-Smith R., Jones D.~H.~P., 1990a, MNRAS, 
246, 637 

\bibitem[\protect\citeauthoryear{Friend et al.}{1990}]{1990MNRAS.246..654F} 
Friend M.~T., Martin J.~S., Connon-Smith R., Jones D.~H.~P., 1990b, MNRAS, 
246, 654 

\bibitem[Goodchild \& Ogilvie(2006)]{2006MNRAS.368.1123G} Goodchild, S., \& 
Ogilvie, G.\ 2006, MNRAS, 368, 1123 

\bibitem[Harrison et al.(2004)]{2004AJ....127..460H} Harrison, T.~E., 
Johnson, J.~J., McArthur, B.~E., Benedict, G.~F., Szkody, P., Howell, 
S.~B., \& Gelino, D.~M.\ 2004, AJ, 127, 460 

\bibitem[Hauschildt et al.(1999)]{1999ApJ...512..377H} Hauschildt, P.~H., 
Allard, F., \& Baron, E.\ 1999, ApJ, 512, 377

\bibitem[Henry \& McCarthy(1993)]{1993AJ....106..773H} Henry, T.~J., \& 
McCarthy, D.~W., Jr.\ 1993, AJ, 106, 773 

\bibitem[Kirkpatrick(2005)]{2005ARA&A..43..195K} Kirkpatrick, J.~D.\ 2005, 
ARA\&A, 43, 195 

\bibitem[Kolb(1993)]{1993A&A...271..149K} Kolb, U.\ 1993, A\&A, 271,
149 

\bibitem[Kolb \& Baraffe(1999)]{1999MNRAS.309.1034K} Kolb, U., \& Baraffe, 
I.\ 1999, MNRAS, 309, 1034 

\bibitem[Kolb \& Baraffe(2000)]{2000NewAR..44...99K} Kolb, U., \& Baraffe, 
I.\ 2000, New Astronomy Review, 44, 99 

\bibitem[Kolb et al.(2001)]{2001MNRAS.321..544K} Kolb, U., King, A.~R., \& 
Baraffe, I.\ 2001, MNRAS, 321, 544 

\bibitem[Leggett et al.(1996)]{1996ApJS..104..117L} Leggett, S.~K., Allard, 
F., Berriman, G., Dahn, C.~C., \& Hauschildt, P.~H.\ 1996, ApJS, 104,
117 

\bibitem[Leggett et al.(2001)]{2001ApJ...548..908L} Leggett, S.~K., Allard, 
F., Geballe, T.~R., Hauschildt, P.~H., \& Schweitzer, A.\ 2001, ApJ, 548, 
908 

\bibitem[McArthur et al.(1999)]{1999ApJ...520L..59M} McArthur, B.~E., et 
al.\ 1999, ApJl, 520, L59 

\bibitem[McArthur et al.(2001)]{2001ApJ...560..907M} McArthur, B.~E., et 
al.\ 2001, ApJ, 560, 907 

\bibitem[Mennickent et al.(2004)]{2004MNRAS.347.1180M} Mennickent, R.~E., 
Diaz, M.~P., \& Tappert, C.\ 2004, MNRAS, 347, 1180 

\bibitem[Paczy{\'n}ski(1971)]{1971ARA&A...9..183P} Paczy{\'n}ski, B.\ 1971, 
ARA\&A, 9, 183 

%\bibitem[Patterson (1998)]{1998PASP..110.1132P} Patterson J., 1998,
%PASP, 110, 1132  

\bibitem[Patterson et al.(2003)]{2003PASP..115.1308P} Patterson, J., et 
al.\ 2003, PASP, 115, 1308 

\bibitem[Patterson et al.(2005)]{2005PASP..117.1204P} Patterson, J., et 
al.\ 2005, PASP, 117, 1204 


\bibitem[\protect\citeauthoryear{Pearson}{2006}]{2006MNRAS.371..235P}
Pearson K.~J., 2006, MNRAS, 371, 235


\bibitem[Podsiadlowski et al.(2003)]{2003MNRAS.340.1214P} Podsiadlowski, 
P., Han, Z., \& Rappaport, S.\ 2003, MNRAS, 340, 1214 

\bibitem[Ramseyer(1994)]{1994ApJ...425..243R} Ramseyer, T.~F.\ 1994, ApJ, 
425, 243 

\bibitem[Rappaport et al.(1983)]{1983ApJ...275..713R} Rappaport, S., 
Verbunt, F., \& Joss, P.~C.\ 1983, ApJ, 275, 713 

\bibitem[Rieke \& Lebofsky(1985)]{1985ApJ...288..618R} Rieke, G.~H., \& 
Lebofsky, M.~J.\ 1985, ApJ, 288, 618 

\bibitem[Ritter \& Kolb(2003)]{2003A&A...404..301R} Ritter, H., \& Kolb, 
U.\ 2003, A\&A, 404, 301 

\bibitem[Smith \& Dhillon(1998)]{1998MNRAS.301..767S} Smith, D.~A., \& 
Dhillon, V.~S.\ 1998, MNRAS, 301, 767 

\bibitem[Stehle et al.(1996)]{1996MNRAS.279..581S} Stehle, R., Ritter, H., 
\& Kolb, U.\ 1996, MNRAS, 279, 581 

\bibitem[Stump et al. (2002)]{} Stump, D et al. 2002, Phys. Rev. D,
  65, 014012

\bibitem[Szkody et al.(2004)]{2004AJ....128.1882S} Szkody, P., et al.\ 
2004, AJ, 128, 1882 

\bibitem[Thorstensen et al.(2002)]{2002PASP..114.1117T} Thorstensen, J.~R., 
Fenton, W.~H., Patterson, J., Kemp, J., Halpern, J., \& Baraffe, I.\ 2002a, 
PASP, 114, 1117 

\bibitem[Thorstensen et al.(2002)]{2002ApJ...567L..49T} Thorstensen, J.~R., 
Fenton, W.~H., Patterson, J.~O., Kemp, J., Krajci, T., \& Baraffe, I.\ 
2002b, ApJl, 567, L49 

\bibitem[Townsley \& Bildsten(2003)]{2003ApJ...596L.227T} Townsley, D.~M., 
\& Bildsten, L.\ 2003, ApJl, 596, L227 

\bibitem[Verbunt(1987)]{1987A&AS...71..339V} Verbunt, F.\ 1987, A\&AS, 71, 
339 

\bibitem[Warner(1995)]{1995cvs..book.....W} Warner, B.\ 1995,
'Cataclysmic Variable Stars', Cambridge  Astrophysics Series,
Cambridge, New York: Cambridge University Press

\bibitem[Thorstensen(2003)]{2003AJ....126.3017T} Thorstensen, J.~R.\ 2003, 
AJ, 126, 3017 

\bibitem[Patterson(1984)]{1984ApJS...54..443P} Patterson, J.\ 1984, ApJS, 
54, 443 

\bibitem[Webbink(1990)]{1990apcb.conf..177W} Webbink, R.~F.\ 1990, in 
Accretion-Powered Compact Binaries, Ed: C. W. Mauche, Cambridge:
Cambridge University Press, p177,

\bibitem[Webbink \& Wickramasinghe(2002)]{2002MNRAS.335....1W} Webbink, 
R.~F., \& Wickramasinghe, D.~T.\ 2002, MNRAS, 335, 1 

\end{thebibliography}
\end{document}